\documentclass[aip,graphicx]{revtex4-1}
\usepackage{graphicx}
\draft % marks overfull lines with a black rule on the right

\begin{document}

%\preprint{}

\title{Subcritical transition to turbulence of a precessing flow in a cylindrical vessel} %Title of paper



\author{Johann Herault} 
 \email{j.herault@hzdr.de}
 \author{Thomas Gundrum}
 \author{ Andre Giesecke}
 \author{ and  Frank Stefani}
\affiliation{Helmholtz-Zentrum Dresden-Rossendorf, P. O. Box 510119,D-01314 Dresden, Germany.}


\date{\today}

\begin{abstract}
The transition to turbulence in a precessing cylindrical vessel is experimentally investigated. Our measurements are performed for a {  nearly-resonant} configuration with an initially laminar flow dominated by an inertial mode with  azimuthal wave number $m=1$ superimposed on a solid body rotation. By increasing the precession ratio, we observe a transition from the laminar to a non-linear regime, which then breakdowns to turbulence   for larger precession ratio.  Our measurements show that the transition to turbulence is subcritical, with a discontinuity of the wall-pressure and  the power consumption at the threshold $\epsilon_{LT}$. The turbulence is self-sustained below this threshold, describing a bifurcation diagram with a hysteresis. In this range of the control parameters, the turbulent flows can suddenly collapse after a finite duration, leading to a definitive relaminarization of the flow. The average lifetime  $\langle \tau \rangle$ of the turbulence increases rapidly when $\epsilon$ tends to $\epsilon_{LT}$.% d isplay a either a critical behaviour below the threshold $\epsilon_{LT}$;%, with  $\langle \tau \rangle \sim \exp( a \vert \epsilon-\epsilon_{LT} \vert^{-\beta} )$ ($a, \beta>0$) {    or an exponential law}.  
\end{abstract}


\maketitle %\maketitle must follow title, authors, abstract and \pacs

\section{Introduction}
%--------------------------------------------------------------------------------------------
Rotating flows are ubiquitous in geophysical, astrophysical and industrial context. When an external torque is applied perpendicularly to the rotation axis of a solid body, it  displays  a precessing motion  in order to conserve the angular momentum. The precession refers to the continuous change of direction of the axis of rotation, which rotates around the axis of precession. The response of a fluid flow driven by the precession of the vessel is a complex process\cite{le2014flows}, which results mostly in the interplay of inertial waves, Ekman boundary layers  and base flow. It has been observed experimentally that the fluid flow driven by precession can amplify \cite{malkus1968precession} a magnetic field, and simulations have demonstrated that a dynamo effect  can operate \cite{tilgner2005precession,nore2011nonlinear}. The present study belongs to the framework of the project DRESDYN\cite{stefani2012,stefani2014,Giesecke2014}, which aims to investigate experimentally a  precession driven dynamo.

The observed   regimes and instabilities of a precessing flow depend strongly on the configuration, which includes the geometry (spherical, ellipsoidal or cylindrical), the rotation of the body, and the strength of the precession. For sake of clarity, we recall only the results  in  cylindrical geometry. In most of the experimental investigation, the set-up is composed of a cylinder rotating around its revolution axis at the rotation rate $\Omega_c$ and this cylinder  is mounted on a turntable rotating at the precession  rate $\Omega_t$. When the angle between the axes of rotation and precession is non-zero, the flow deviates from  the  solid body rotation, due to the   acceleration of the rotation axis. Initially observed by McEwan \cite{mcewan1970inertial},  the laminar flow is composed  of Kelvin modes superimposed on a solid body rotation\cite{manasseh1994distortions,kobine1995inertial,meunier2008rotating,liao2012flow}. The Kelvin modes \cite{Kelvin1880} (or inertial modes) are the eigen-modes of a rotating cylindrical vessel.  Each mode has its own eigen-frequency $\omega$, which depends on the radial, axial and azimuthal wave numbers. The gyroscopic force due to the precession excites the  modes  with azimuthal wave number $m=1$ and  angular frequency $\Omega_c$.  A mode can be resonant if its eigen-frequency   is close  to the angular forcing frequency $\Omega_c$. Out of the resonance, the amplitude of the modes scales like $\vert \omega- \Omega_c \vert^{-1}$ and it increases with the strength of the precession, quantified by the precession ratio $\epsilon=\Omega_t / \Omega_c$. At the resonance, viscous and non-linear effects play an important role in the saturation of the amplitude of the Kelvin modes \cite{kudlick,kobine1995inertial,kerswell1995viscous,meunier2008rotating}.

Suggested by McEwan\cite{mcewan1970inertial}, Kerswell \cite{kerswell1999secondary} has demonstrated that the Kelvin modes forced by   precession become  unstable by a  parametric instability. The mechanism is based on a triadic resonance with two other free Kelvin modes, i.e. two modes  not directly forced by the precession. Lagrange \textit{et al}. \cite{lagrange2008instability,lagrange2011precessional} confirmed experimentally this scenario for a precessing flow with a single forced Kelvin mode close to its resonance. The parametric instability can either saturate \cite{lagrange2011precessional} or trigger a secondary instability   called resonant collapse\cite{mcewan1970inertial,manasseh1992breakdown,lin2014experimental}. A weakly non-linear analysis performed by Lagrange \textit{et al} \cite{lagrange2011precessional} predicts a subcritical bifurcation leading to a stationary state or a cyclic dynamics for larger Reynolds number. When the instability does not saturate, a sequence of bifurcations called resonant collapse triggers a breakdown of the flow into a  small-scale chaotic flow. Manasseh \cite{manasseh1992breakdown} pointed out that the most violent resonant collapse occurs  for the resonant Kelvin mode with the smallest radial wave number (called   type $A$ breakdown), leading to a fine-scale turbulence.  The resonant collapse can display a cyclic dynamics with a transient chaotic phase  followed by a relaminarization of the flow \cite{mcewan1970inertial,lin2014experimental}.  During the chaotic phase,  the  kinetic energy remains mostly in modes with  frequencies in the   range of those of the inertial modes\cite{lin2014experimental}.

It is tempting to speculate that the turbulence  appears  progressively from the chaotic set associated with the resonant collapse, but the experimental observations reports a more complex road to turbulence. Indeed, by  further increasing  the precession ratio, Lin \textit{et al}  \cite{lin2014experimental}  observed  a non-linear regime after the resonant collapse, characterized by the disappearance of the free Kelvin modes in the power-spectrum.  At  critical precession ratio,  a sustained intense turbulent flow fills the vessel \cite{mouhali2012evidence,gans1971hydromagnetic}.  Gans \cite{gans1971hydromagnetic} reported a discontinuity of  the   torque  curve  at the transition to turbulence. The torque curves exhibits also a hysteresis between the turbulent and non-turbulent branches. A similar bistability   has been reported by Mouhali \textit{et al} \cite{mouhali2012evidence} in a cylinder and  by Malkus\cite{malkus1968precession} in a ellipsoidal geometry. The presence  of a discontinuity and the hysteresis suggests that a subcritical instability  triggers the turbulence for these experiments with  large Reynolds number ($Re >10^5$).   The nature of the transition to turbulence remains unclear.

In the framework of dynamical systems, the so-called subcriticality refers to a transition from a base state   to a new state (here turbulent), such that the new state is created from one  or a sequence of bifurcations \cite{Manneville}, which does not change the local stability of the base state. It is analogue to a first-order phase transition, like the  vapour-liquid transition of   water  below the critical point. The transition is generally abrupt and violent and the triggering depends on the strength of the perturbation if the flow remains linearly stable. In the past fifty years, great progress  has been made to understand the subcritical bifurcation to turbulence in shear flows, like pipe flows \cite{eckhardt2007turbulence} or Plane-Couette flows\cite{daviaud1992subcritical,Dauchot1995,Bottin1998}. Recently, those concepts have  been extended to magnetohydrodynamics flows \cite{Ponty2007,herault2011,riols2013}. Subcritical bifurcation in shear flows is characterized by specific properties \cite{grossmann2000onset,eckhardt2007turbulence}, such as   non-normal growth, complex structures of the flow pattern and finite lifetimes of the turbulence \cite{hof2006finite,bottin1998statistical,borrero2010transient,rempel2010supertransient,grants2003experimental}. 

The finite lifetime  of the turbulence is one of the most  striking features of shear flows undergoing a subcritical bifurcation to turbulence. The perturbations leading to turbulent spots may relaminarize after a finite duration. The  lifetimes of turbulence are random and their mean value increases rapidly with the Reynolds numbers. It turns out that the turbulence is a metastable  regime for a  range of Reynolds numbers. This feature has been observed in hydrodynamic shear  flows (pipe flow \cite{hof2006finite}, Plane-Couette flow\cite{bottin1998statistical} or stable Taylor-Couette flow \cite{borrero2010transient}) and   in magnetohydrodynamics shear-flows \cite{rempel2010supertransient,grants2003experimental}.

In the present study, we investigate the subcritical transition to turbulence in a fluid flow driven by a precessing  cylindrical vessel.  The observed transition to turbulence is a robust process  and occurs above a well defined threshold  different from the threshold of the parametric instability of the forced Kelvin mode.  Below the threshold of the subcritical transition, the turbulence is self-sustained, describing an hysteresis bifurcation diagram. We show that the turbulence  is   metastable in the hysteresis region and can cease suddenly.

Section \ref{experiment} introduces the experimental set-up and the dimensionless control parameters. In section \ref{physical_reg}, the regimes appearing before the turbulence are briefly reviewed. The quantitative features of the turbulence and the associated subcritical bifurcation are presented in section \ref{subcritical}. Section \ref{section_life_time} presents the experimental investigation of the finite lifetime of the turbulence.

% Indeed,  

%--------------------------------------------------------------------------------------------

%-------------------------------------------------------------------------------------------
\section{The Experimental device}
\label{experiment}

\subsection{Experimental setup}

\begin{figure}[ht]
\begin{center}
 \includegraphics[width=45mm,height=45mm]{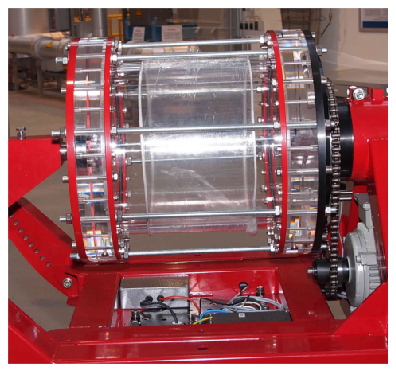}
  \includegraphics[width=45mm,height=45mm]{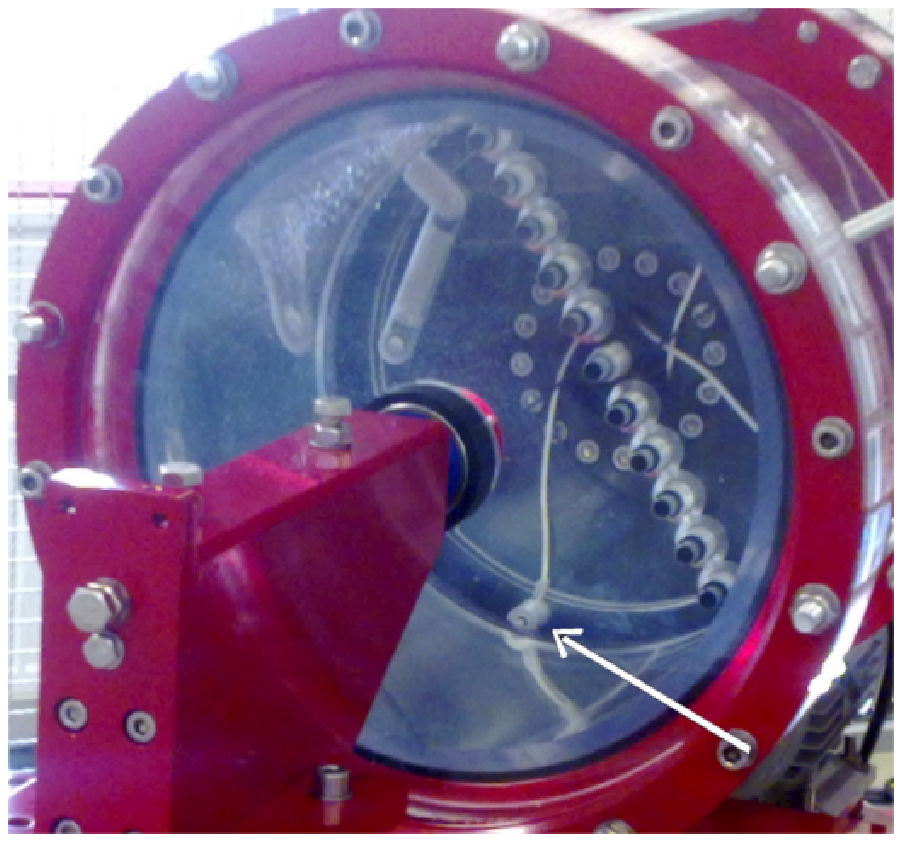}
\includegraphics[width=55mm,height=45mm]{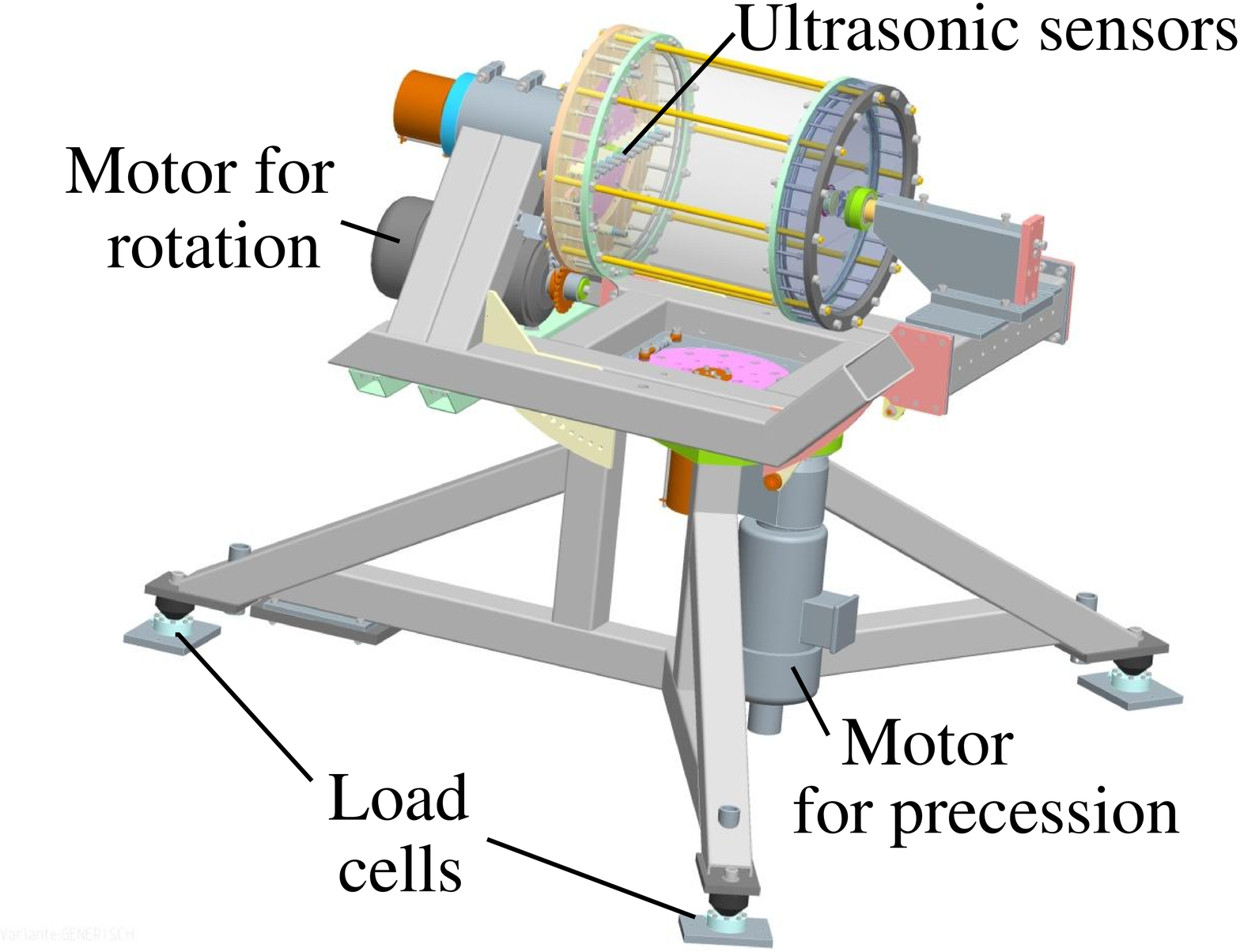}
 
\caption{Left: photography of the precessing cylinder. Center: transversal photography. The  white pressure sensor is located at the bottom of the end-cap, close to the lateral-wall (white arrow). An array of UDV sensors is also visible. Right : sketch of the experiment.}
\label{fig_setup}
\end{center}
\end{figure}

The experimental setup is illustrated on the Fig. \ref{fig_setup}. The   vessel is a cylinder of radius $R=163$ mm and height $h=326$ mm. It is completely filled with water. For the visualization of the flow (see Fig.\ref{fig_regimes}), a small amount of air is introduced. {    The bubbles are only used to visualize the flow pattern and  all the quantitative results have been obtained with a very small amount of bubbles}. The container rotates around its symmetry axis. The cylinder is driven   by   an asynchronous $3$ kW  motor via a transmission chain.  The power is     supplied by a slip ring. To improve the parallelism of the end caps of the cylinder, they are joined together by eight rods (yellow rods on the right Fig.\ref{fig_setup}). The end-cap located at the opposite side of the  motor is maintained by a counter bearing, in order to avoid any bending. This structure is mounted  on the turntable. It can be tilted to vary the inclination between the direction of  the rotation of the cylinder and the direction of the precession. In the present study, the angle is constant at 90${}^{\tiny o}$ so that rotation axis and precession axis are orthogonal like in refs \cite{lin2014experimental,mouhali2012evidence}.  The turntable is driven by a second asynchronous $2.2$ kW motor.  The frequency rate of the turntable $f_t$ can be increased up to $1$Hz.  {    We have not used a control loop for the motors, in order to avoid perturbations due to the control process, like self excitations, which stress the mechanics of
the experiment.} The rotation rates of  the cylinder and the turntable   are continuously measured by two tachometers. The fluctuations of the rotation rates are smaller than $0.05 \%$ and {    the rotation rates do not vary during the transitions from and to the turbulent regime.}

%\vspace{6cm}

%--------------------------------------------------------------------------------------------------
\subsection{Measurement techniques}

\subsubsection{Power consumption measurement}

The power consumption  of the three-phase asynchronous motors  is calculated by measuring the input current on a single phase and the voltage difference between two phases. Here, we report   only the power consumption of the motor of the cylinder. The measurement of the current is performed with a Tektronix current-clamp TCP312 and amplified by a TCPA300 amplifier with a typical accuracy of $1\%$. The voltage is measured  by a differential probe Sapphire SI-9010. The amplitudes of the current $I(t)$,  the amplitude of the  voltage difference  $U(t)$ and   their relative  phase $\phi(t)$  are extracted by a Hilbert transform. The internal resistance of the motor $R_I$   has been measured by a 4 wires technique after a long run, in order to measure $R_I$ close to the operating conditions (effect of the internal heating). The functions $U(t)$, $I(t)$ and $\phi(t)$ vary on time scales much larger than the rotation period of the motor. The instantaneous  power consumption $P(t)$ of an equilibrated motor is defined by

\begin{equation}
P(t)= \frac{\sqrt{3}}{2} U(t) I(t) \cos \left( \phi(t)+\frac{\pi}{6}  \right)-\frac{3}{4}R_I I(t)^2.
\end{equation}

\noindent  Because $I(t)$ and $U(t)$ are the peak  values, the product $U  I $ and $I^2$ must be multiplied by the factor $1/2$ to obtain the RMS values. The relative phase $\phi(t)$ has been shifted by $\pi/6$ in order to obtain the relative phase between the current and the voltage for the same phase.  The average power consumption is given by $P_m= T^{-1} \int^T _0 P \hbox{dt}$, with $T$ the duration of the measurement. The value $P_m$  can be decomposed into the sum of the mechanical loss, the magnetic loss and the power dissipated by the  flow. Only the latter is of interest to us. We expect that the magnetic and mechanical losses do not change significantly for constant rotation rate $\Omega_c$. Their values have been measured for runs at  $\Omega_t=0$ (no precession), which  corresponds to a solid body rotation without  power dissipated by the  flow. Our measurements show that the power consumption increases like $\Omega_c^2$ for $\Omega_t=0$.

\subsubsection{Pressure measurement}

The pressure is measured at the end-cap close to the motor (see center Fig.\ref{fig_setup}). Two XPM5 miniature pressure sensors are mounted flush on the rotating end-cap. They are on the same radius with respect to the axis of revolution, located at   $r=160$ mm. Their diameter   is $3.6$ mm. In the present study, we focus only on the pressure variation. In   section \ref{physical_reg}, we present  the frequency power spectrum. The power spectra are rescaled in order to have  a power spectral density $PSD[p](f)$ tending to $1$ for low frequencies. In the rest of the paper (section \ref{section_life_time} and \ref{subcritical}), we have used a smoothing filter  to low-pass filter the signal with a typical cut-off frequency $f_c/2$.   The purpose of the filter is to suppress the components with frequencies equal  or  larger than the frequency rate of the cylinder $f_c$ in order to diagnose the variation of the slow components.

%--------------------------------------------------------------------------------------------------

\subsection{Dimensionless number}

We recall all the parameters characterizing our set-up in the    table I.

\begin{table}[htb!]
\label{tab1}

 \begin{center}
 
\begin{tabular}{|ll ll l|}
\hline
\rule[-1ex]{0pt}{4ex} \textbf{Parameters} && \textbf{Definition}& & \textbf{Value}    \\
\hline
\rule[-1ex]{0pt}{4ex}  $\nu$ & \hspace{2cm}&  Kinematic viscosity & \hspace{2cm} & $10^{-6} $m${}^2$  s${}^{-1}$   \\
\rule[-1ex]{0pt}{4ex}  $R$ &&  Radius && $163$ mm \\
\rule[-1ex]{0pt}{4ex} $h$ &&  height && $326$ mm  \\
\rule[-1ex]{0pt}{4ex} $ f_c= \Omega_c / 2 \pi$ &&  Rotation frequency && $0-10$Hz  \\
\rule[-1ex]{0pt}{4ex} $f_t= \Omega_t / 2 \pi$ &&  Precession frequency && $0-1$Hz  \\
\rule[-1ex]{0pt}{4ex} $\Theta $ &&  Precession angle && $90{}^{\hbox{\tiny{o}}}$  \\
\hline
\end{tabular}
 
\vspace{0.15cm}

\caption{Physical parameters of the experimental setup.}   
\end{center}
\end{table}

The dimensional analysis allows us to define three dimensionless numbers from the five parameters $(h,R,\nu,\Omega_c,\Omega_t)$ (we consider $\Theta$ as a dimensionless number)

\begin{equation}
\Gamma=\frac{h}{2R}, \quad \quad \epsilon=\frac{\Omega_t}{\Omega_c} , \quad \quad \quad Re=\frac{\Omega_c R^2}{\nu}
\end{equation}

\noindent $\Gamma$ is the aspect ratio and is equal to $1$. It plays an important role in the determination of the eigen-frequency of the Kelvin modes \cite{Greespan1969}. The Reynolds numbers $Re$ quantifies the dissipative process and can be varied in the range $[1.6-16.5] \times 10^5$. The effects of the precession are quantified by the precession ratio $\epsilon$, which varies in the range $0-0.16$. Only the latter two dimensionless numbers can be modified. In the present study, the Reynolds number is  only changed  for studying the scaling of the power dissipated at constant $\epsilon$ (section \ref{Hysteresis}) and the parameter space (section \ref{Effects_Reynolds}). In the most part  of the paper, the Reynolds number is fixed  to $Re=5.65 \times 10^5$.

%--------------------------------------------------------------------------------------------

%---------------------------------------------------------------------------------------------------
\section{The flow regimes }
\label{physical_reg}

\begin{figure}[ht]
\begin{center}
\includegraphics[width=160mm,height=43mm]{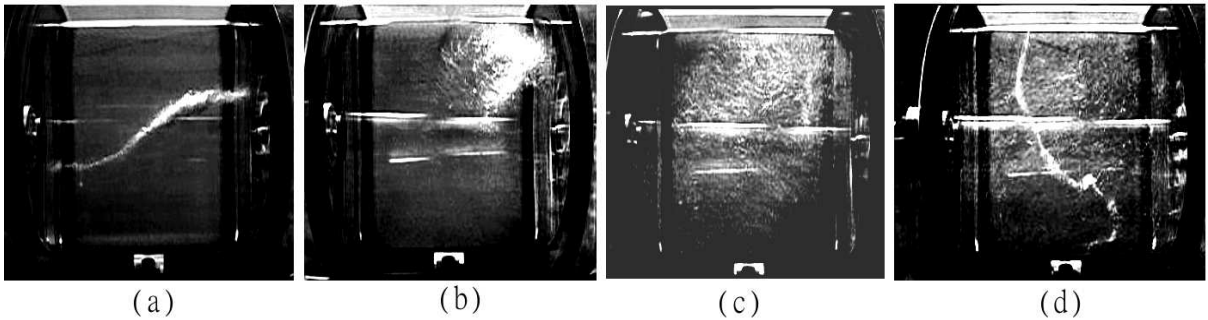}
 
\caption{The different regimes of the precessing flow are observable, via a small amount of bubbles,  in the precessing reference frame (turntable frame) for $f_c=10Hz$ and $\epsilon=0.74\times 10^{-2}$ (a), $5.9\times 10^{-2}$ (b), $7.4\times 10^{-2}$ (c), and  $10\times 10^{-2}$ (d). {    The amount of air is kept constant}. From left to right : standing forced Kelvin mode with a S-shape (a),  non-linear regime with almost not distinguishable  Kelvin modes and {    the bubbles localized in the upper left region}  (b), turbulent regime with an important spreading of the bubbles (c), strong vortex superimposed with turbulent fluctuations (d). }
\label{fig_regimes}
\end{center}
\end{figure}

\begin{figure}[ht!]
\begin{center}
\includegraphics[width=80mm,height=50mm]{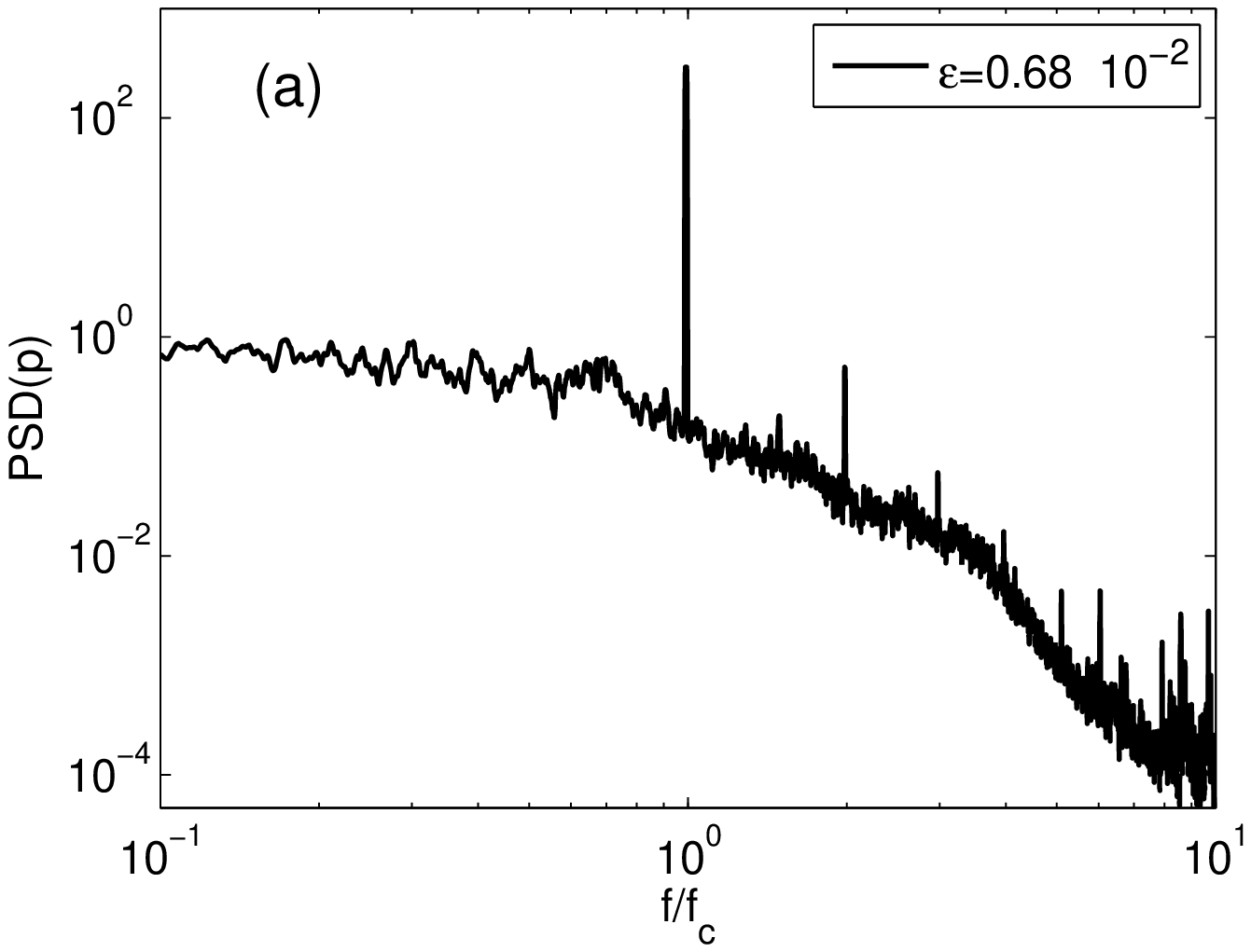}
\includegraphics[width=80mm,height=50mm]{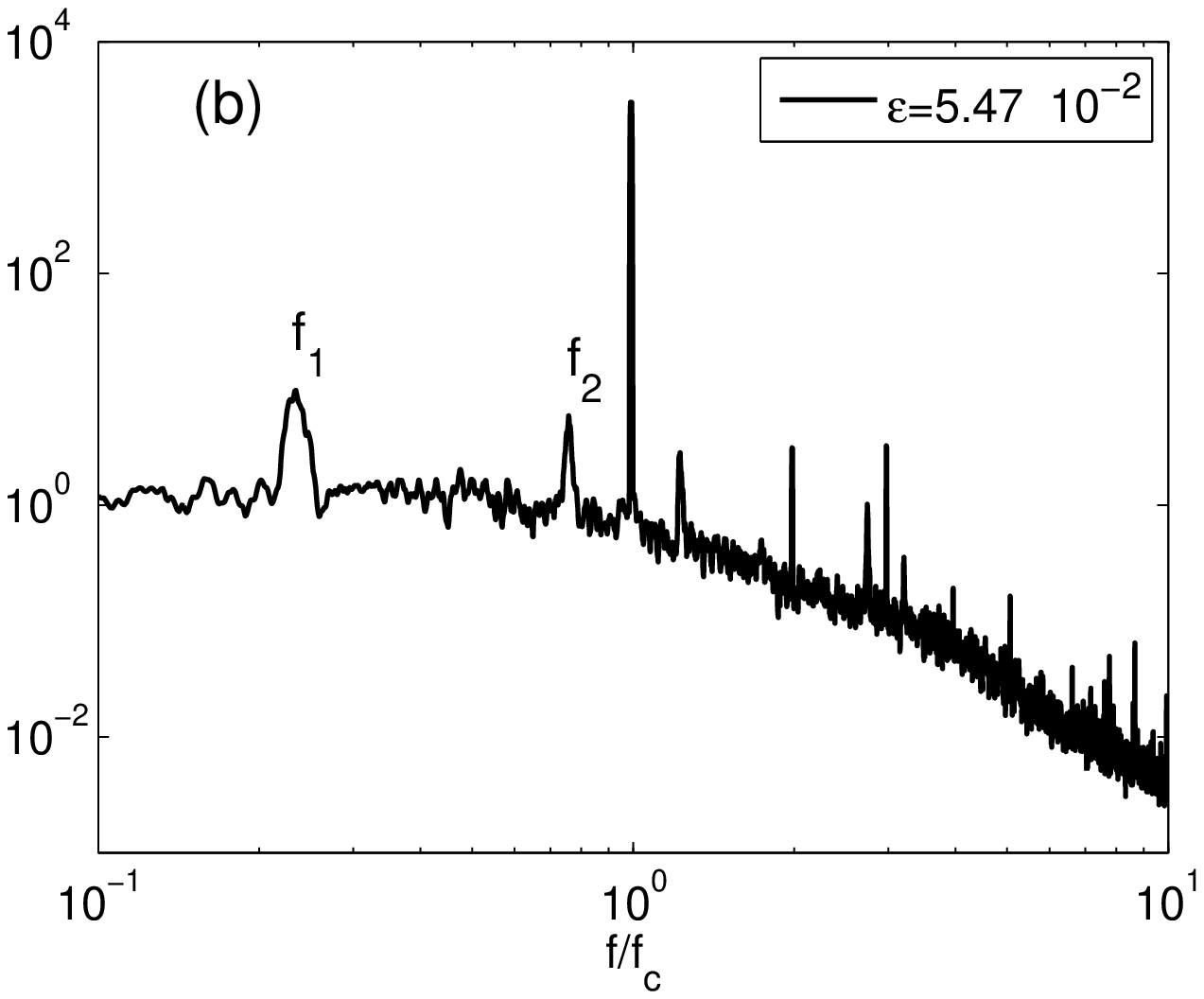}

\includegraphics[width=80mm,height=50mm]{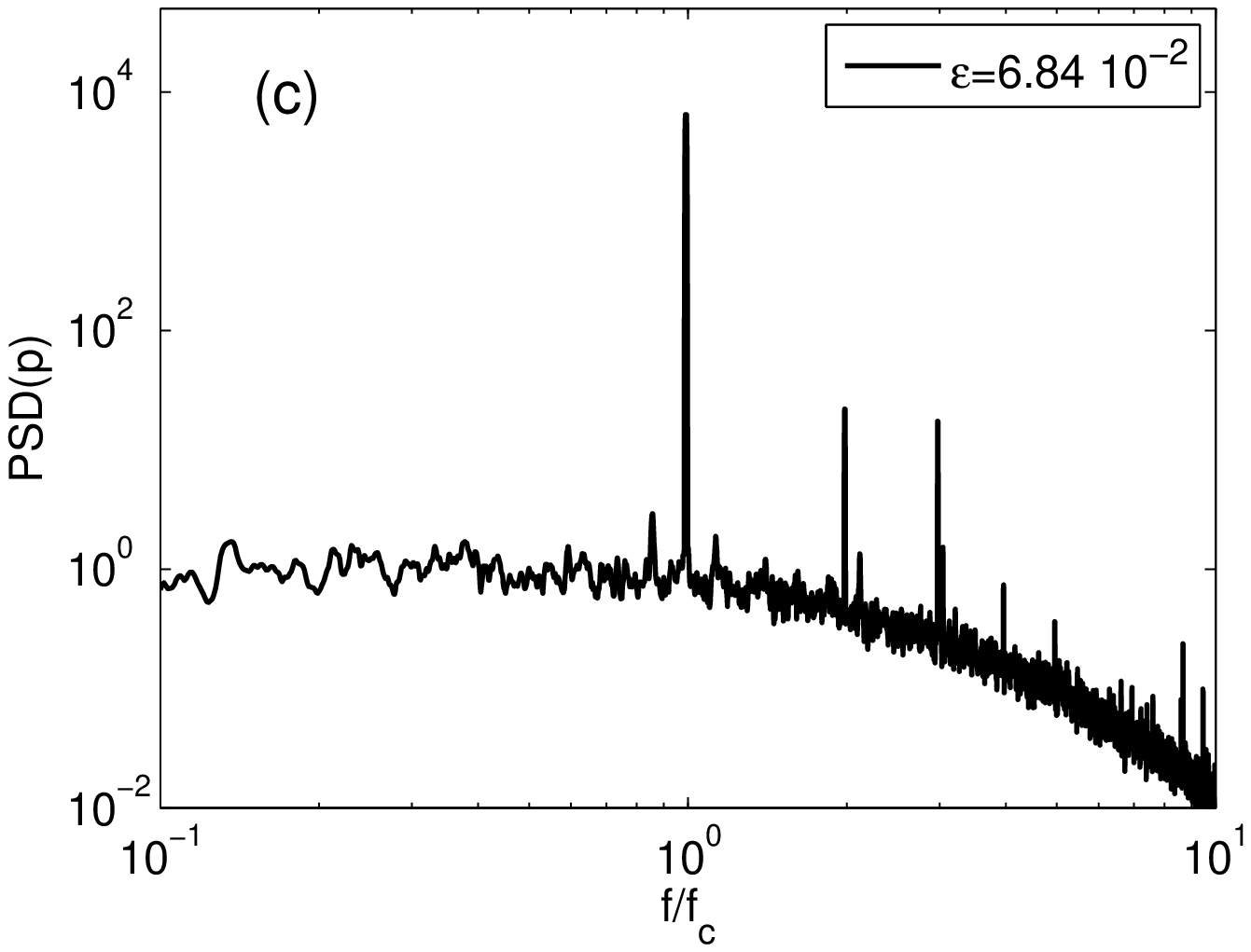}
\includegraphics[width=80mm,height=50mm]{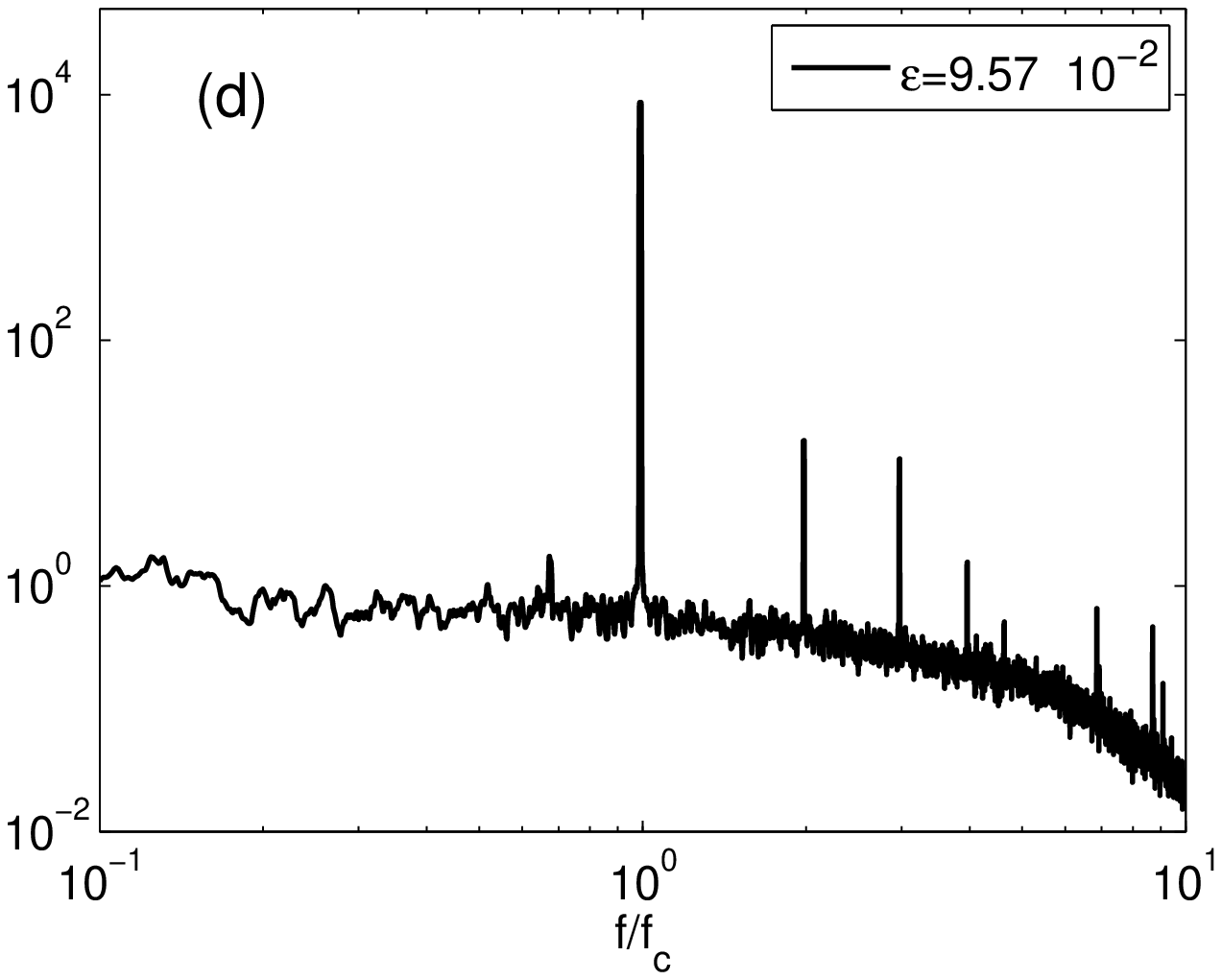}
 
\caption{Power spectrum of the pressure signal for   $Re=5.65 \times 10^5$  and $\epsilon=0.68 \times 10^{-2}$(b) (laminar regime), $\epsilon=5.47 \times 10^{-2}$(b) (non-linear regime with triadic resonance), $\epsilon=6.84 \times 10^{-2}$(c) (non-linear regime without triadic resonance) and $\epsilon=9.57 \times 10^{-2}$ (d)(turbulent regime).}
\label{spectre_eps}
\end{center}
\end{figure}

Before focusing on the transition to turbulence, we present the different regimes observed in the precessing experiment. We   distinguish three  regimes by increasing progressively the precession ratio $\epsilon$ for a constant Reynolds number $Re$, thanks to a direct visualization of the flows (see Fig.\ref{fig_regimes})  or  by analysing the power spectrum of the pressure signal  (see Fig.\ref{spectre_eps}). Visualization and pressure measurements have been performed for two different Reynolds numbers: $Re=14\times 10^5$ for the visualization and $Re=5.65 \times 10^5$ for the pressure measurements.

Our classification  holds for  the range $Re \in[1,14] \times 10^5$ and the parametric study of the regimes is reported in the section \ref{Effects_Reynolds} (cf  Fig. \ref{parameter_space}). The  three   regimes are

{ \centering  \begin{enumerate}
\item the laminar regime,
\item the non-linear regime,
\item the turbulent regime.
\end{enumerate}  \centering
} 
  
The laminar flow corresponds to a superposition of Kelvin modes with an azimuthal wave number  $m=1$. Following the theory of precessing flow  \cite{Greespan1969}, the first resonance occurs at $f=0.996 f_c$ for an aspect ratio $h=1$. Our configuration corresponds to a resonant case with one dominant  Kelvin mode. In the Fig.  \ref{fig_regimes}(a), we clearly identify the S-shape of the minimal pressure region, associated with the axial wave number  $k_z=\pi \cdot L^{-1}$. It is standing in the turntable frame. The power spectrum of the pressure signal exhibits an intense peak at the frequency $f_c$ and at the corresponding harmonics (see Fig.\ref{spectre_eps}(a)). {    The peaks at $2f_c$ and $3f_c$ likely arise from non-linear self-interaction of the forced Kelvin mode \cite{meunier2008rotating}. For a forced Kelvin mode defined by the frequency $f_c$ , an azimuthal wave number $m=1$ and an axial wave number $k_0$ , the non-linear interaction will drive the modes ($2f_c$, $m = 2$, $k = 0$) and ($0,0, k = 2k_0$) (the other combinations are forbidden). Moreover, the inertial pressure term, given by $|\Omega\times \vec{r}|^2$, contributes to the component $f = 2f_c$ \cite{meunier2008rotating}}. The centrifugal pressure and the hydrostatic pressure contribute also to the pressure components at $f_c$.  

We define as non-linear regime  the   regime  between the first instability of the Kelvin mode $m=1$ and  the transition to turbulence. At a given precessing ratio, the forced Kelvin mode becomes unstable. Indeed, the peaks observed on the power spectrum (see Fig.\ref{spectre_eps}(b)) suggest the existence of two free Kelvin modes   $(f_1=0.223 f_c$ and $f_2=0.75 f_c$) in near-resonance with the forced Kelvin mode ($f_1+f_2=0.97f_c$). The thresholds of appearance of these triads will   be discussed later (Fig. \ref{parameter_space}). {    At the threshold, the peaks in the power-spectrum correspond to frequencies close to $f_1=0.35f_c$ and $f_2=0.65f_c$, which may be assigned to $m = 5$ and $m = 6$ modes. By further increase of the precession ratio, these frequencies vary. We believe that this effect is due to the saturation process which is achieved by a detuning in frequency of the modes (see for instance \cite{lagrange2011precessional}). These results will be reported in a future paper.}

Further increasing the precession ratio, we have identified a sequence of bifurcations, and we are currently designing   a velocimetry set-up   to diagnose in details these bifurcations. These results will be reported in a future paper. For larger precession ratio, the two previous frequency peaks disappear (see Fig.\ref{spectre_eps}(c)) and the power-spectrum becomes flatter. The forced Kelvin mode is  no more distinguishable by direct visualization (see Fig.\ref{fig_regimes}(b)).{     The bubbles are localized close to one end-cap due to the breaking of the centro-symmetry of the laminar flow.  }

These observations suggest that the dynamics is already complex and {    likely} chaotic. Our measurements are   consistent with the   observations of Lin \textit{ et al} \cite{lin2014experimental}, who  observed the disappearance  of triadic resonance of inertial modes in the strongly non-linear regime. 

Further increasing the precession ratio, the flow becomes turbulent. The properties of this transition will be detailed in the following sections. Visually, the turbulent flow is characterized by the spreading of small bubbles in the full vessel (see Fig.\ref{fig_regimes}(c)). Occasionally, we observe a vortex filament aligned with the direction of   precession (see Fig.\ref{fig_regimes}(d)). The pressure signal gives few information  about the turbulent flow (see Fig. \ref{spectre_eps}(d)). The power-spectrum becomes almost flat on the range of frequencies $f<4 f_c$. 
%--------------------------------------------------------------------------------------------

\section{Turbulence and subcritical bifurcation}
\label{subcritical}

In this section, we focus on the transition to turbulence. First, we confirm the existence of a discontinuity and a hysteresis of the power consumption at the transition \cite{gans1971hydromagnetic}. Then, we study the dynamics and the robustness of this transition. We conclude this section by the systematic determination of the threshold in function of the Reynolds number. 
%--------------------------------------------------------------------------------------------
\subsection{Power consumption: hysteresis and scaling}
\label{Hysteresis} 

\begin{figure} [htb!]
\begin{center}
\includegraphics[width=100mm,height=60mm]{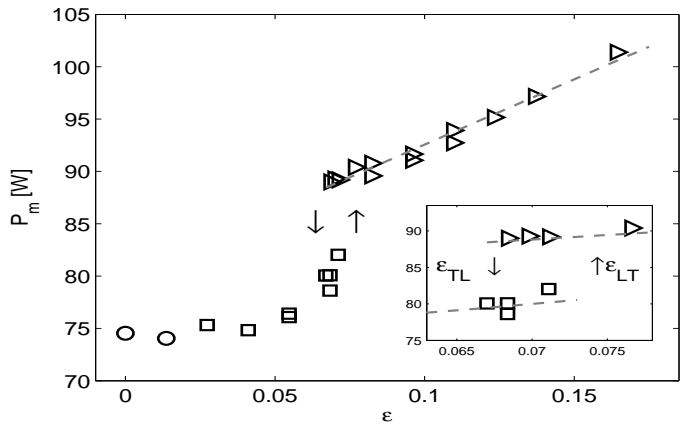}  
\caption{Mechanical power consumption of the motor of the rotating cylinder for $\Omega_c=(8 \pi)$ rad/s and $\epsilon=\Omega_t/\Omega_c \in [0,0.165]$. Two distinct branches are clearly observable:  the lower branch ($L$) with the laminar regime (circle) and  the non-linear regimes  (square), and the turbulent branch ($T$) (triangle). Dashed line: linear fit  with  $P_m=P_{m0}+C \epsilon$. Insert: zoom on the hysteresis. The system jumps from the $L$ branch to the $T$ branch at $\epsilon_{LT}=(7.3 \pm 0.1 )\times 10^{-2}$ and goes back from $T$ to $L$ at $\epsilon_{TL}=(6.7 \pm 0.1) \times 10^{-2}$. }
\label{fig_powerc}
\end{center}
\end{figure}

We carried out measurements of power consumption at constant Reynolds number $Re=5.65 \times 10^5$ ($f_c=4$Hz).  We have performed a first ascending series of measurements from $\epsilon=0$ to $16.4\times 10^{-2}$ and   a second descending series from $10 \times 10^{-2}$ to $6.56 \times 10^{-2}$. Due to the presence of fluctuations at low frequencies (see Fig.\ref{spectre_eps}),  each measurement  lasts $40$ minutes ($9600$ periods) in order to converge the average power consumption  $P_m(\epsilon)$.  The results are reported on Fig. \ref{fig_powerc}.

The average power consumption $P_m(\epsilon)$  displays two distinct branches: the  lower branch $L$ (squares and circles) with $\epsilon \in [0,\epsilon_{LT}]$ and the turbulent branch $T$ (triangles) with $\epsilon \in [ \epsilon_{TL}, 16\times 10^{-2}]$. The $L$ branch corresponds to the  laminar regime (circle) and to the non-linear regime (square). Both branches are disconnected and the system undergoes a subcritical bifurcation from $L$ to $T$ at $\epsilon_{LT}$ and goes back from $T$ to $L$ at $\epsilon_{TL}$. The thresholds are  $\epsilon_{LT}=7.3 \pm 0.1 \times 10^{-2}$ and $\epsilon_{TL}=6.7 \pm 0.1 \times 10^{-2}$ and the   hysteresis is clearly visible with $\epsilon_{LT} -\epsilon_{TL} \simeq 6 \times 10^{-3}$ (insert, Fig.\ref{fig_powerc}). The determination of $\epsilon_{LT}$ is very robust (section \ref{transition_turb}) whereas a precise measurement of $\epsilon_{TL}$ is very difficult (section \ref{metastability}).

The behaviour of the power consumption  of the $L$ and $T$ branches are quite different.  On the $L$ branch, the power consumption  does not  vary significantly for $\epsilon<0.05$ and increases  rapidly between $\epsilon \simeq 0.05$ and $\epsilon_{LT}$. In this range, the resonant triads  disappear   (section \ref{physical_reg}).  The   power consumption on the $T$ branch exhibits a  linear growth  (dashed grey line on Fig.\ref{fig_powerc}) with

\begin{equation}
P_m (\epsilon  ) = P_{ 0}+C  \epsilon    
\label{eqDIss}
\end{equation}
 
\noindent with $ P_{0}=80 $W and $C=124 W$. A similar behaviour  was observed in the torque   measurement of Gans \cite{gans1971hydromagnetic} for the hydrodynamic case.  
 
\begin{figure} [htb!]
\begin{center} 
\includegraphics[width=90mm,height=60mm]{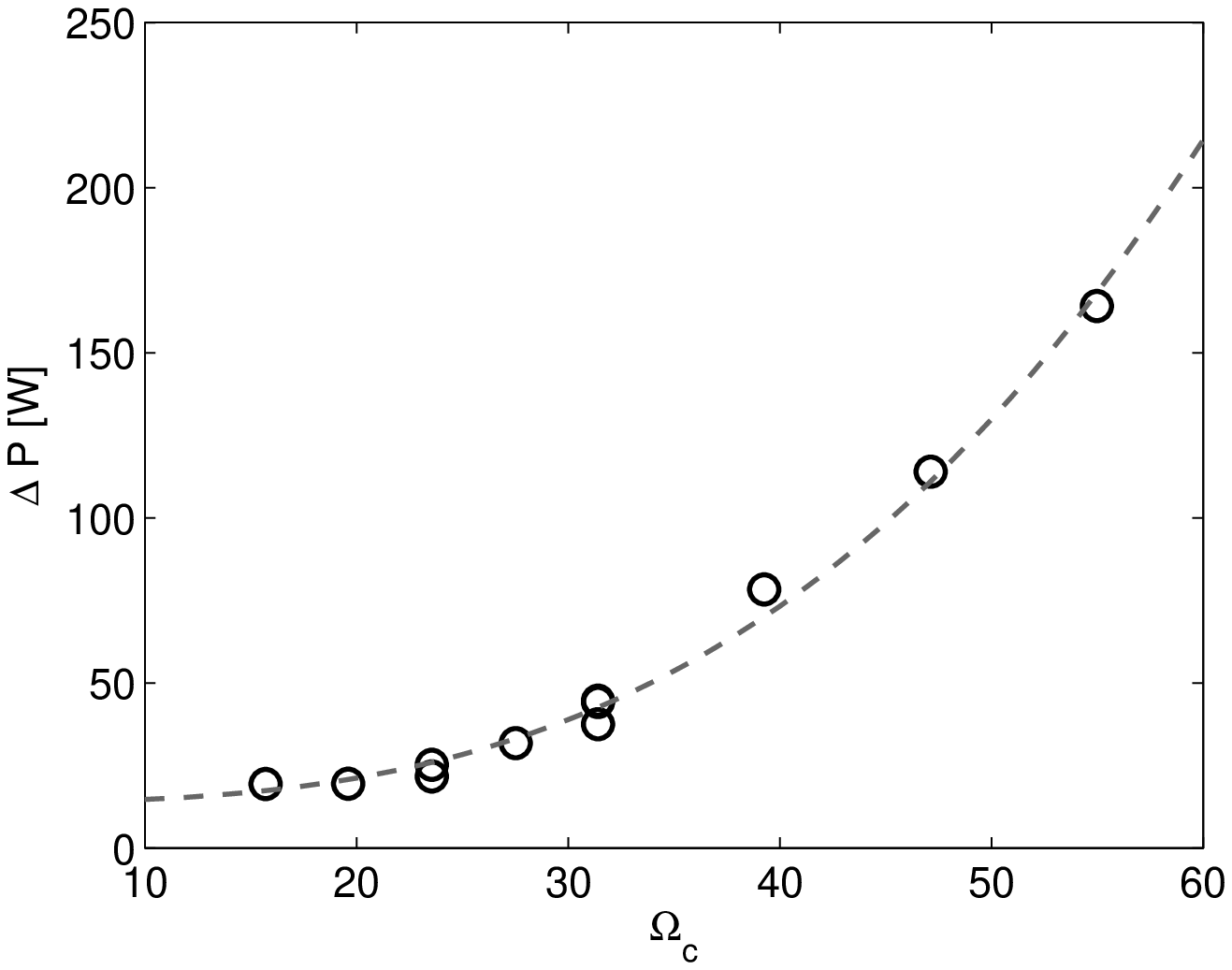} 
\includegraphics[width=70mm,height=60mm]{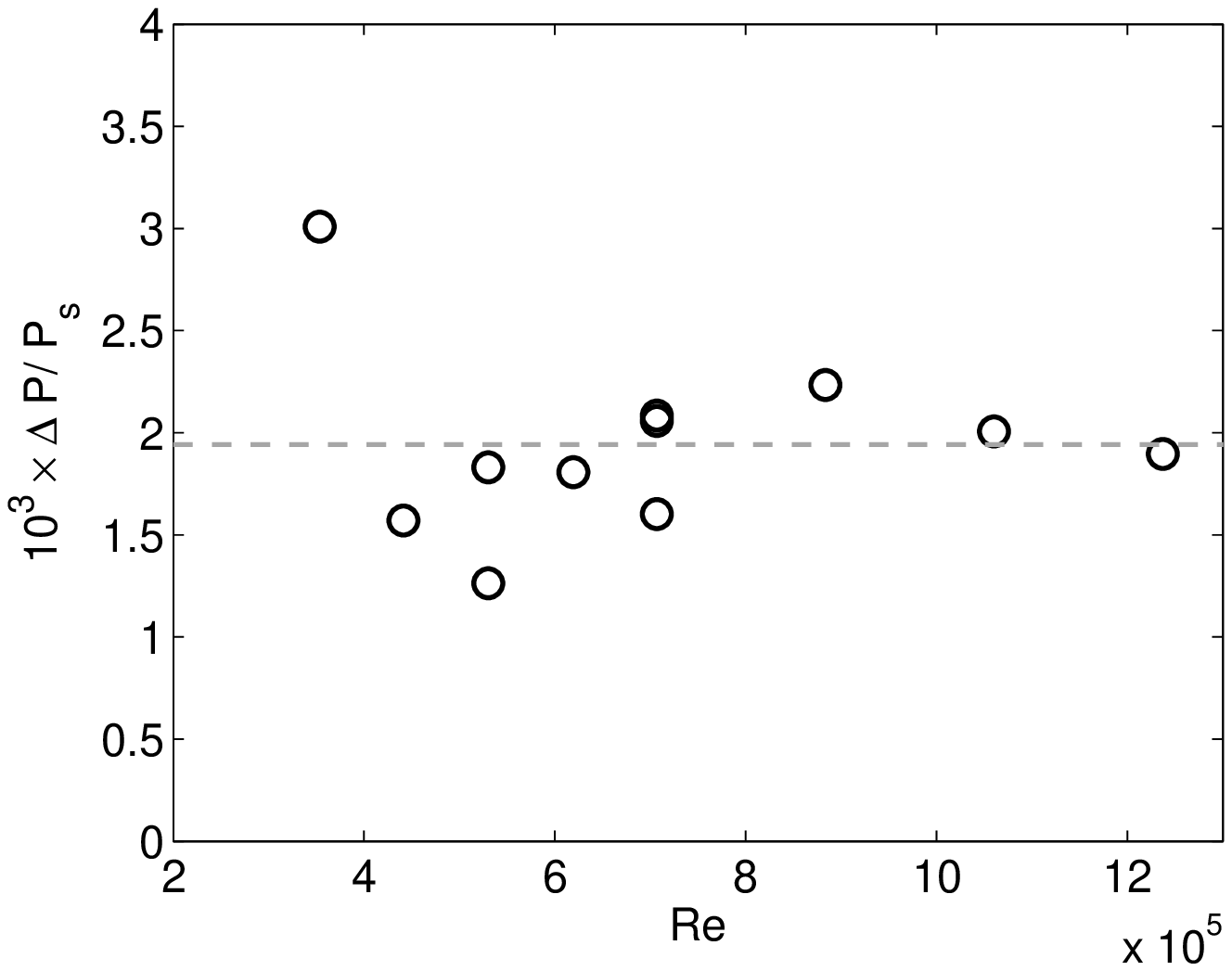} 
\caption{Left: power injection, $\Delta P=P_m(\epsilon)-P_m(0)$ for $\epsilon= 8.7 \times 10^{-2}$ and $Re \in [3.5,12] \times 10^5$. The dashed grey curve follows the power-law $\Delta P\sim \Omega_c^3$. Right: compensated power $\Delta P/P_s$ with $P_s= \rho 2 \pi R^4 L \Omega_c^3$.}
\label{fig_powerc2}
\end{center}
\end{figure}

We performed another series  of measurements with constant $\epsilon=8.75\times 10^{↨-2}$ in order to estimate the Reynolds dependence of the power consumption by varying $\Omega_c$. The flow remains turbulent for the associated Reynolds numbers, which vary in the range   $ [3.5,13] \times 10^5$. To compare the measurements, we subtract the power consumption at $\epsilon=0$ to remove the mechanical loss (increasing like $\Omega_c^2$) of the set-up.  The power injected in the flow  $\Delta P=P_m(\epsilon)-P_m(0)$ is reported on the left hand side of Fig. \ref{fig_powerc2}. It  increases like $ \Omega_c^3$ for   constant $\epsilon$. The dimensionless power $\Delta P/P_s$, with $P_s= \rho 2 \pi R^4 L \Omega_c^3$, is roughly constant for $Re \in [3.5,13] \times 10^5$ (right Fig.\ref{fig_powerc2}), suggesting that the dimensionless injected power  becomes independent of the Reynolds number. 

From equation \ref{eqDIss}  and the behaviour shown in Fig. \ref{fig_powerc2}, it can be concluded that the power dissipated into the turbulent flow scales like

\begin{equation}
\frac{\Delta P}{ \rho   R^4 L \Omega_c^3} \sim   \epsilon    
\label{eqDIss2}
\end{equation}
 
This scaling differs from the linear viscous theory with $\Delta P/P_s \sim  \epsilon^2 Re$ and the theoretical constant bounds, $\Delta P/P_s \sim \epsilon^0$, expected for large Reynolds numbers and an infinite cylinder \cite{kerswell1996upper}.  {    Additional information about the turbulent flow is required to
explain the $\epsilon$-dependence of the above scaling law by dimensional analysis.}
%-----------------------------------------------------------------------------

\subsection{Transition to the turbulence}
\label{transition_turb}

\begin{figure} [htb!]
\begin{center} 
\includegraphics[width=95mm,height=60mm]{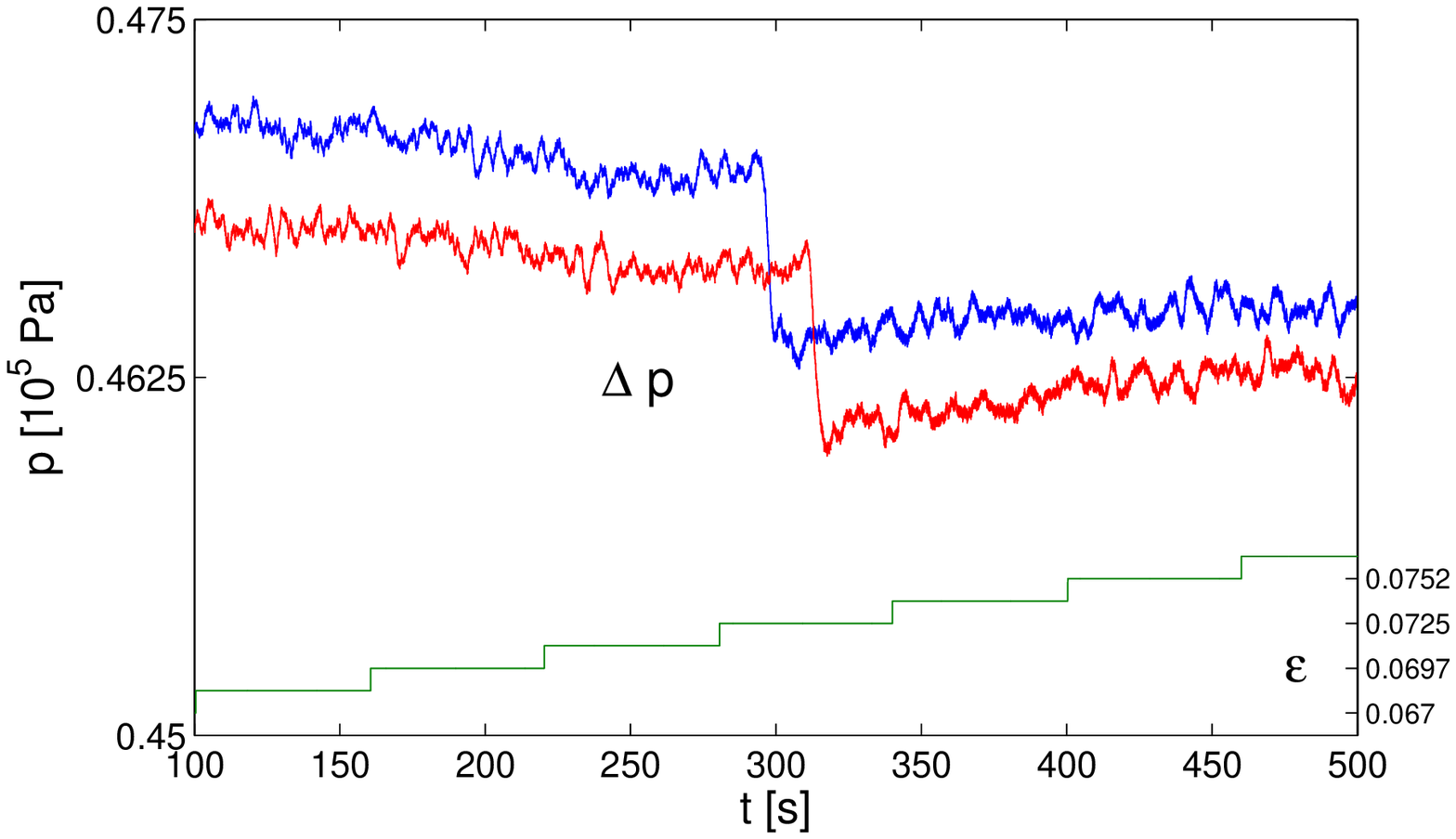}  
\includegraphics[width=65mm,height=60mm]{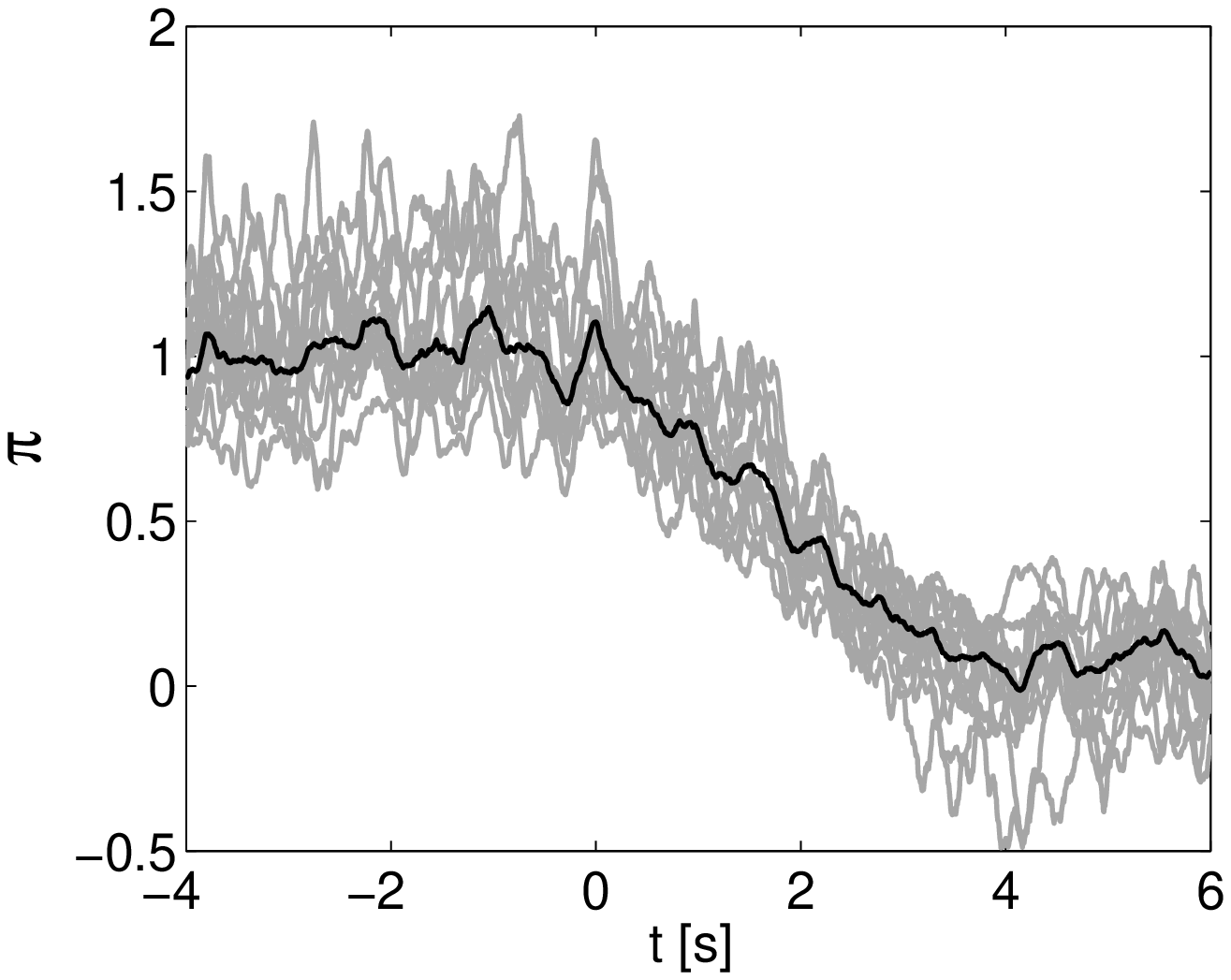} 
\caption{Left: evolution of the low-pass filtered  pressure signal $p$ (blue and red) for two different realizations, and precession ratio $\epsilon$ (black) as a function of   time $t$. Right: rescaled pressure $\pi$ (equation \ref{rescale_press}) during the transition. The different realizations are in grey, and the averaged realization is in black.}
\label{fig_onset}
\end{center}
\end{figure}

We have performed a series of measurements to investigate the transition to turbulence at the threshold $\epsilon_{LT}$ at a given Reynolds numbers  $Re=5.65 \times 10^{5}$. The precession ratio is increased step by step with a plateau time of  60 seconds (240 $f_c^{-1}$) and a precession ratio step   $\Delta \epsilon=1.42 \times 10^{-3}$. We performed $16$ runs with the same procedure, using the pressure signal for the detection of the transition (see left hand side Fig.\ref{fig_onset} ). The transition always occurs when the precession ratio reaches the value $\epsilon=7.25 \times10^{-2}$.   The black curve represents the precession ratio (right ordinate) and the blue curve and the  red curve correspond to the pressure signal from the same probe but for two different runs. The  pressure can slightly drift due to the thermal expansion of the water but we did not observe any correlation between the drift of the pressure and the onset of the instability, so we believe that this effect does not play any significant role in the dynamics of the flow. We observe at $t_d=294$s (resp. $t_d=309$s) for the blue (resp. red) curve a drop of the pressure after the growth of the precession ratio to $\epsilon=7.25 \times 10^{-2}$ at $t_p=280$s. It corresponds to the threshold $\epsilon_{LT}$ for the transition to the turbulence. We have  measured the delay $\Delta T=t_d-t_p$ between the change of $\epsilon$  and the beginning of the instability. The average value of the delay $\Delta T$ is  $15$ seconds (60 $f_c^{-1}$) with a standard deviation of $9$ seconds ($0.6 \times  \Delta T$). The delay $\Delta T$ is comparable to the precession period $f_t^{-1}=3.3$s and  could correspond to the time needed to re-establish the base flow, which then becomes unstable.  In any case, the delay $\Delta T$ is much smaller than the typical transient life time of the turbulence in the hysteresis region (see section \ref{section_life_time}). 

The discontinuity of the pressure   confirms the presence of a subcritical transition. The pressure jump  corresponds roughly to $\Delta p \simeq 500$ Pa and is detected simultaneously on the diametrically opposed probes. The transition seems to be global and affects the low frequency components of the pressure field.  The pressure drop is likely due to a slowdown of the  rotation of the bulk flow. An estimation of this slowdown gives $\Delta f=(\Delta p/ \rho)^{1/2}/  (2 \pi R)   \simeq 0.7 $ Hz, corresponding to $17.5\%$ of the imposed frequency rate $f_c$.  The slowdown of the bulk flow in the turbulent regime has also been reported for  precessing cylinders \cite{mouhali2012evidence} and spheres \cite{goto2007turbulence}.

In order to compare the different realizations,  we define a  rescaled pressure $\pi$ such that 

\begin{equation}
\pi(t)=\frac{p(t)-\langle p_{T} \rangle}{\langle p_{L} \rangle-\langle p_T \rangle}
\label{rescale_press}
\end{equation}

\noindent with $\langle p_{L} \rangle$ (resp. $\langle p_{T} \rangle$) the average pressure before (resp. after) the transition to turbulence. The average pressure $\pi$ decreases from $1$ to $0$.  We have used a coherent average to superimpose the curves. It consists in finding the maximum of the correlation function of two different realizations, in order to remove the delay between them. The instant $t=0 $s corresponds to the beginning of the transition. The different realizations (right hand side Fig. \ref{fig_onset}, grey curves) are concentrated around the average curve (black curve) and do not display any exponential growth or exponential relaxation to the turbulent branch.    We estimate the duration  of the transition   $T_i=4 \pm 0.5$s (16 $f_c^{-1}$).

\subsection{Effects of the Reynolds numbers}
\label{Effects_Reynolds}

\begin{figure}[ht]
\begin{center}
\includegraphics[width=80mm,height=50mm]{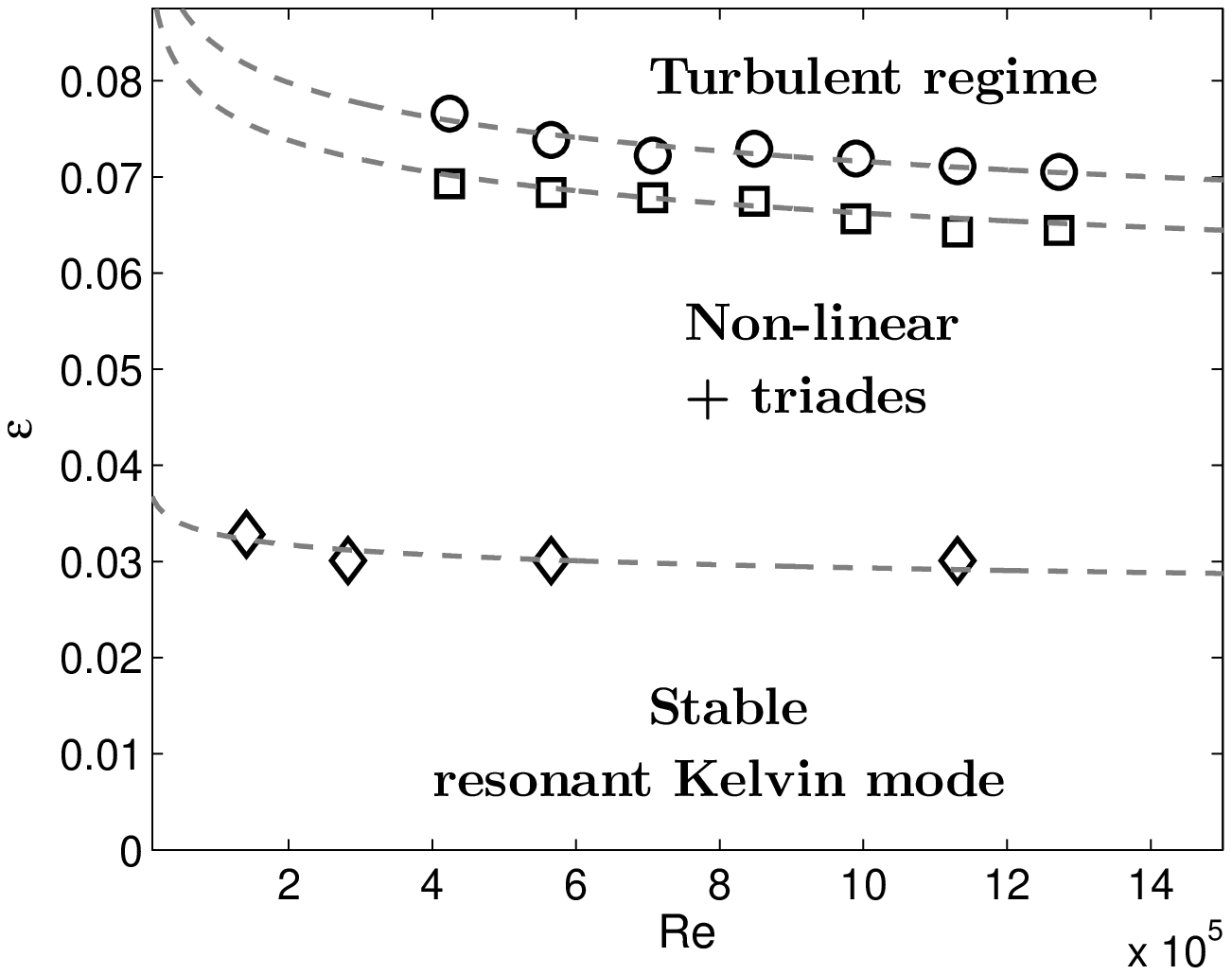}
\includegraphics[width=80mm,height=50mm]{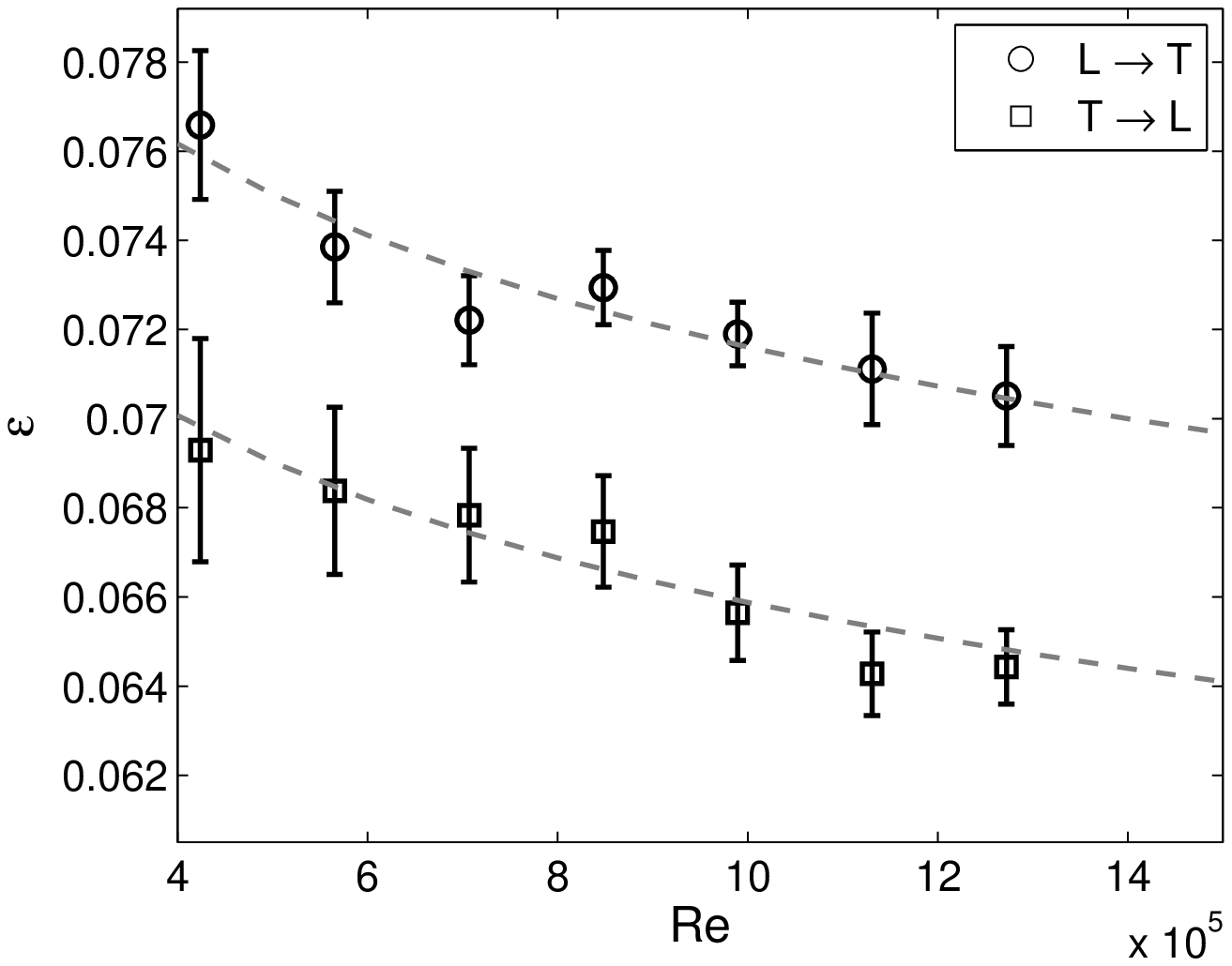}
 
\caption{Left : parameters space $Re$ versus the measured thresholds of the first instability of the forced Kelvin mode (diamonds),  the threshold $\epsilon_{LT}$ of the transition from $L$ branch to $T$ branch  (circles) and $\epsilon_{TL}$ from $T$ to $L$ (squares). Right : zoom on the behaviour of $\epsilon_{LT}$ and $\epsilon_{TL}$. Dashed curves show a fit to a power law with $\epsilon \sim Re^{-0.067}$.}
\label{parameter_space}
\end{center}
\end{figure}

Now, we investigate the dependence  of the threshold as a function of the Reynolds numbers $Re$ . The covered parameter  space $(Re,\epsilon)$ is reported on the left hand side of Fig. \ref{parameter_space}, with $\epsilon_{LT}$ (circle) and $\epsilon_{TL}$ (square). The diamonds represent  the frontier between the laminar flow  and the non-linear regime. It corresponds to the appearance of free Kelvin modes in the power spectrum (section \ref{physical_reg}).  The region between $\epsilon_{LT}$ and $\epsilon_{TL}$ corresponds to the bistable region between the $L$ and the $T$ branch.  The threshold $\epsilon_{LT}$  depends weakly on the Reynolds number, following an empirical power law with  $\epsilon_{LT} \sim Re ^{-0.067}$ (Fig. \ref{parameter_space}, right). The width of the hysteresis  region given by the difference $\epsilon_{LT}-\epsilon_{TL}$, is roughly independent of $Re$.

%--------------------------------------------------------------------------------------------

\section{Finite life time of turbulence}
\label{section_life_time}
%--------------------------------------------------------------------------------------------
\subsection{Metastability}
\label{metastability}

\begin{figure}[htb!]
\begin{center}
\includegraphics[width=70mm,height=50mm]{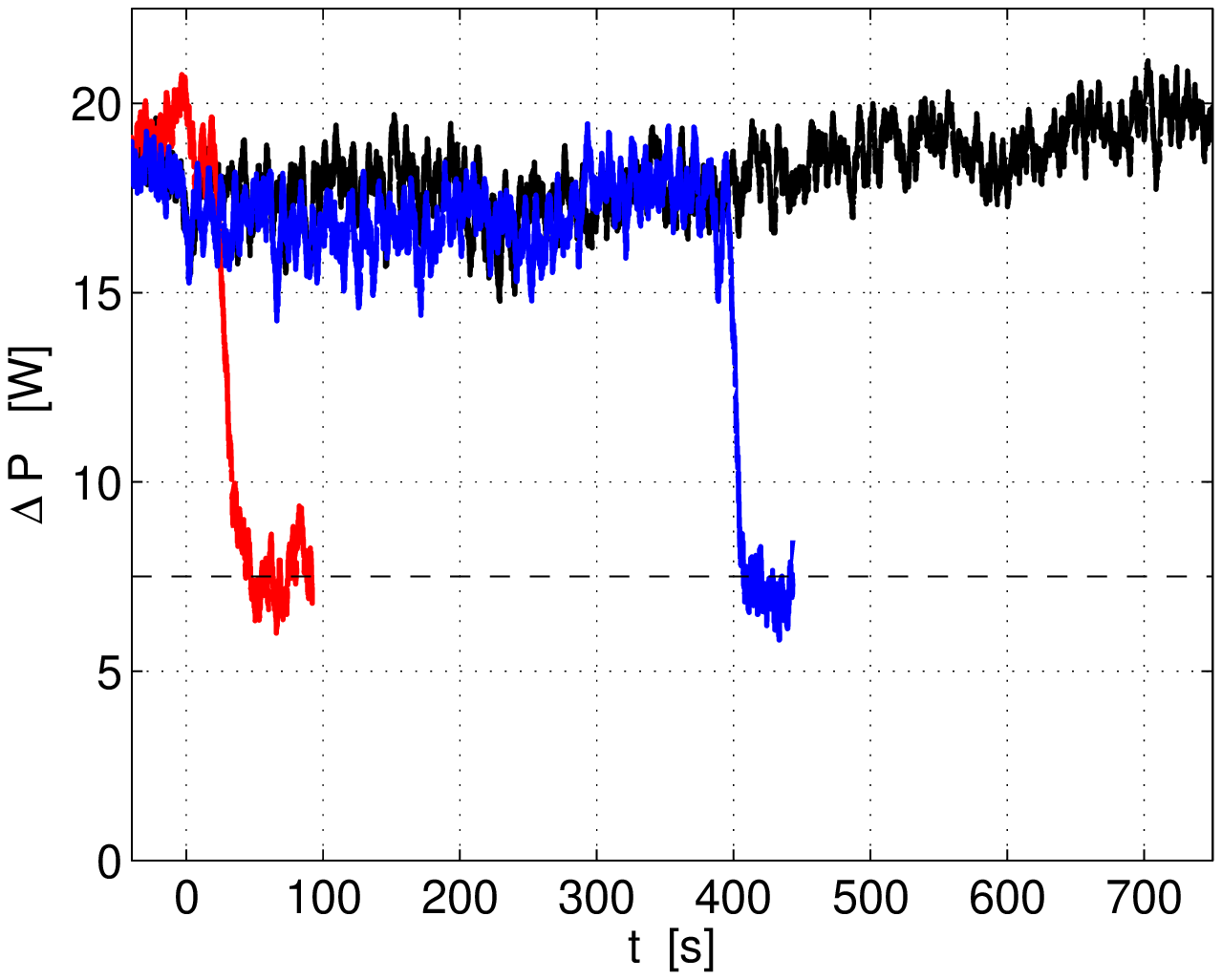} 
\includegraphics[width=70mm,height=50mm]{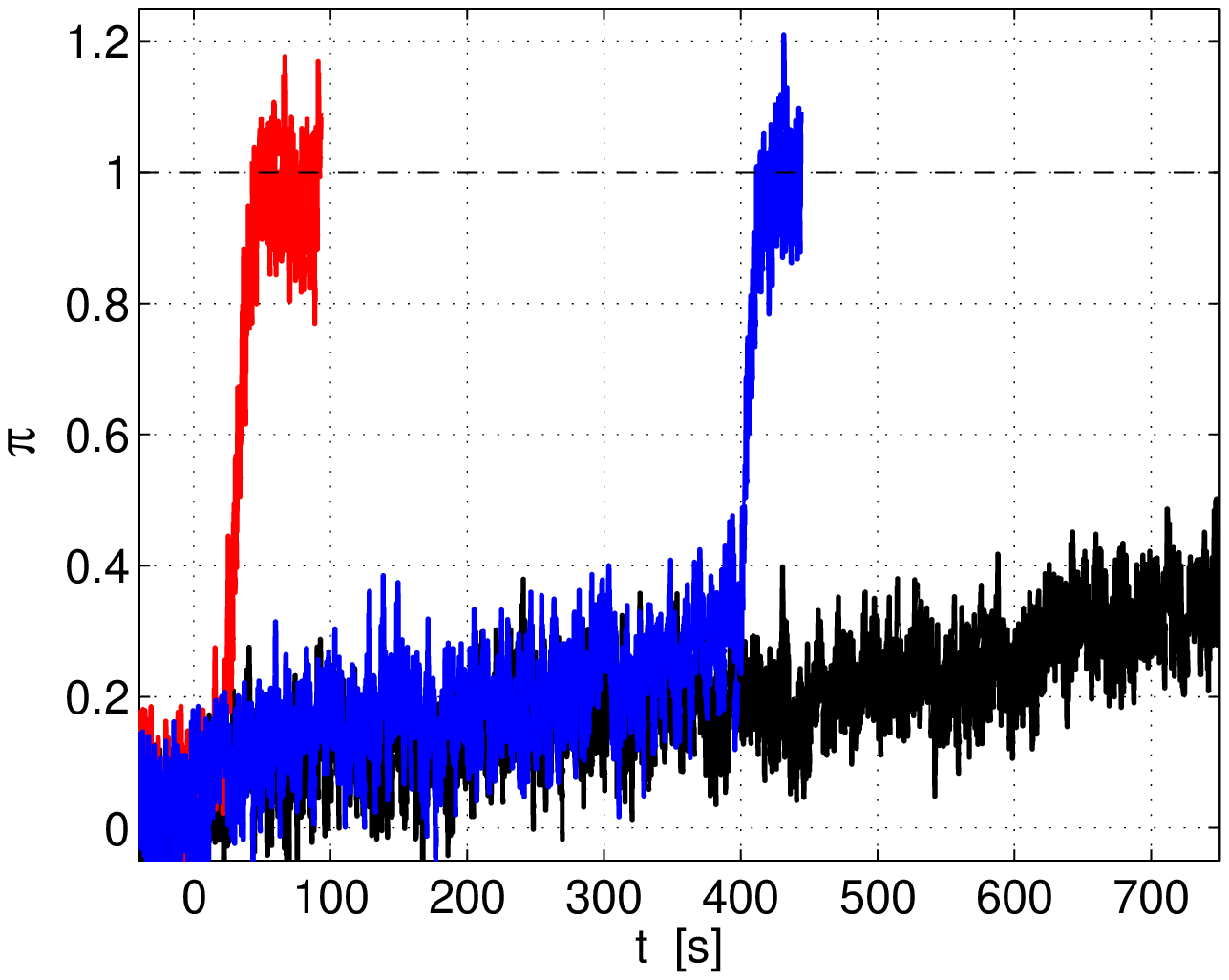} 
\caption{Three different realizations (red, blue, black curves) of the relaxation from the $T$ to $L$ branches for  $Re=5.65 \times 10^{5}$ and $\epsilon=8.5 \times 10^{-2}$ ($>\epsilon_{LT}$) for $t<0$s and  $\epsilon=6.84 \times 10^{-2}$ ($<\epsilon_{LT}$) for $t>0$s. Left: $\Delta P(t)$ power consumption. Right:  rescaled pressure $\pi(t)$. }
\label{decrease_power}
\end{center}
\end{figure}

The turbulent flow exhibits an interesting feature on the range of the hysteresis  $\epsilon \in [\epsilon_{TL},\epsilon_{LT}]$. It can suddenly cease  and relax to the $L$ branch, with a random lifetime in the turbulent branch. Figure \ref{decrease_power}  illustrates this phenomenon. The time series of the power consumption for three different runs  at $Re=5.65 \times 10^{5}$ are displayed on  the left hand side. For $t<0$ s, the flow is turbulent with  $\epsilon=9.3 \times 10^{-2}$ ($>\epsilon_{LT}$) and at $t=0$ s, the rotation rate of the turntable is reduced to $\epsilon=6.84 \times 10^{-2}$ with $ \epsilon_{TL}<\epsilon <\epsilon_{LT}$. The drop of power consumption corresponds to the transition from the $T$ branch to the $L$ branch.  It  occurs rapidly after the slowdown for the red curve,  or $400$s ($1.6 \times 10^{3} f_c^{-1}$) for the blue curve. The third realization (black curve) is still turbulent at $t=700$ s ($2.8 \times 10^{3} f_c^{-1}$). Unlike the resonant collapse \cite{mcewan1970inertial,manasseh1992breakdown}, the relaxation to the $L$ branch is definitive. We did not observe spontaneous transition from the non-linear regime to the turbulent regime below the threshold $\epsilon_{LT}$.

 The pressure measurements display also an abrupt transition  when the turbulence collapses. The corresponding   data  are reported on the  right side of Fig.  \ref{decrease_power}. The rescaled pressure (defined in section \ref{transition_turb}) suddenly increases at the transition. The dynamics of this relaxation is discussed in section \ref{relaxation_sec}. The turbulence seems to be metastable for a given $\epsilon$ in the hysteresis region  $[\epsilon_{TL},\epsilon_{LT}]$. The  lifetimes of the turbulence are random and a statistical study of the lifetimes is performed in the next section. 

As mentioned in the section \ref{subcritical}, the precise determination of $\epsilon_{TL}$ becomes intractable due to the apparent metastability of the hysteresis part of the  turbulent branch.  We carried out  measurements for $\epsilon=6\times 10^{-2}$,i.e.    below $\epsilon_{TL}$, with $(\epsilon-\epsilon_{TL})/\epsilon_{TL} \simeq  10\%$. The transition is almost immediate. The delay lasts  $10$ s ($40 f_c^{-1}$) with a small standard deviation of $1.5$ s. This time scale is comparable to the delay  $\Delta T$ before the beginning of the transition   to turbulence (section \ref{transition_turb}). It turns out that the turbulence is  no more metastable for $\epsilon$ smaller than $\epsilon_{TL}$.

%---------------------------------------------------
\subsection{Lifetime statistics}
%---------------------------------------------------

\subsubsection{Lifetime distribution}

To quantify this stochastic process, we have investigated the statistical properties of the duration of the turbulence in the metastable regime. The experimental procedure consists in starting a run at a precession ratio $\epsilon=9.3\times 10^{-2}$ ($>\epsilon_{LT}=7.25\times 10^{-2}$)  in order to obtain a strong turbulent flow. After $200$s ($8 \times 10^{2} f_c^{-1}$), the rotation rate of the turntable is reduced to the $\epsilon$ wanted. The lifetime $\tau$ is defined as the duration between the end of the slowdown and the pressure rise, corresponding to the relaxation to the $L$ branch (see fig \ref{decrease_power}).  The incertitude concerning the end of the slowdown is of order $\Delta T$ (cf. \ref{subcritical}).

\begin{figure}[htb!]
\begin{center}
  \includegraphics[width=80mm,height=60mm]{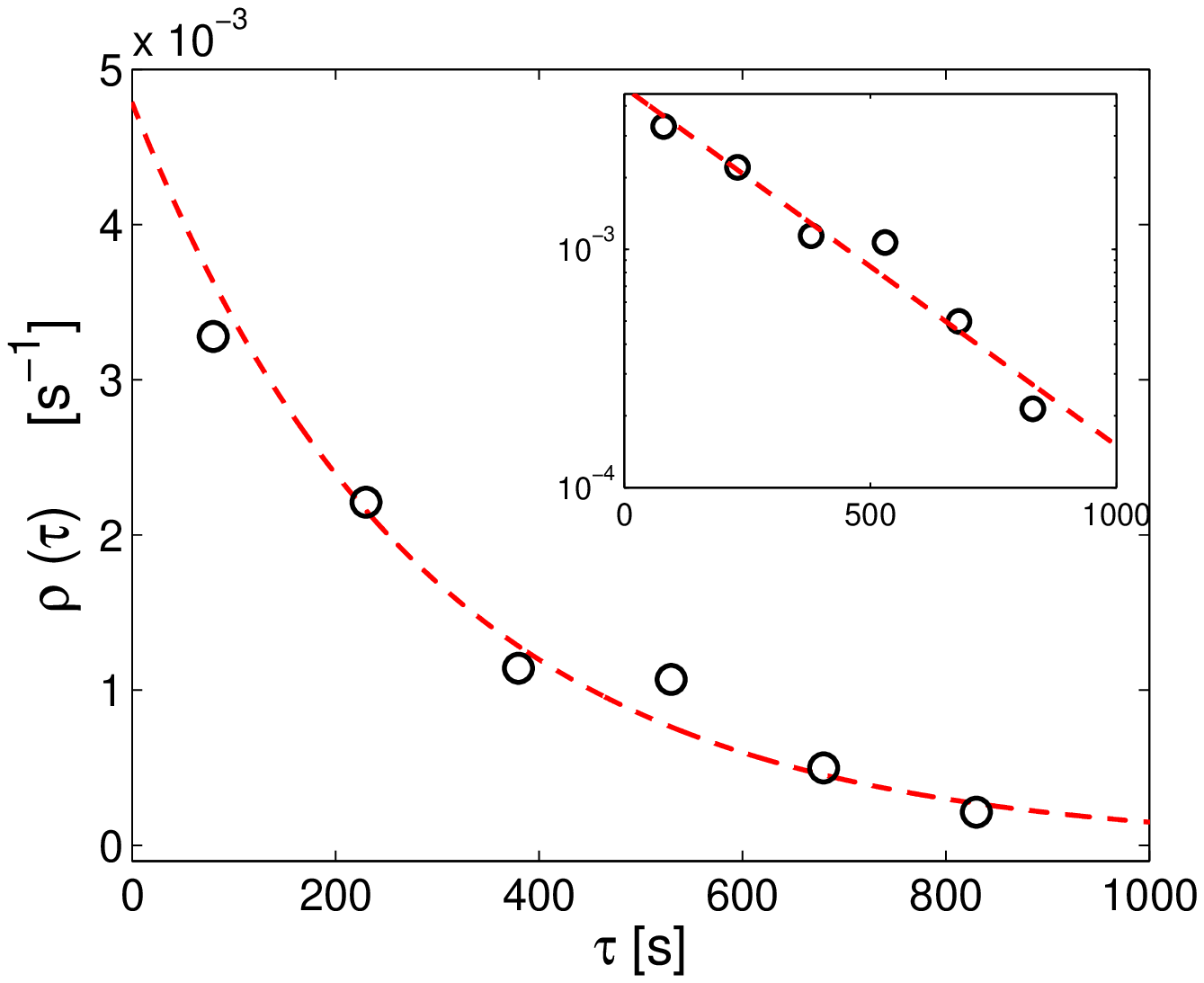}
 \includegraphics[width=80mm,height=60mm]{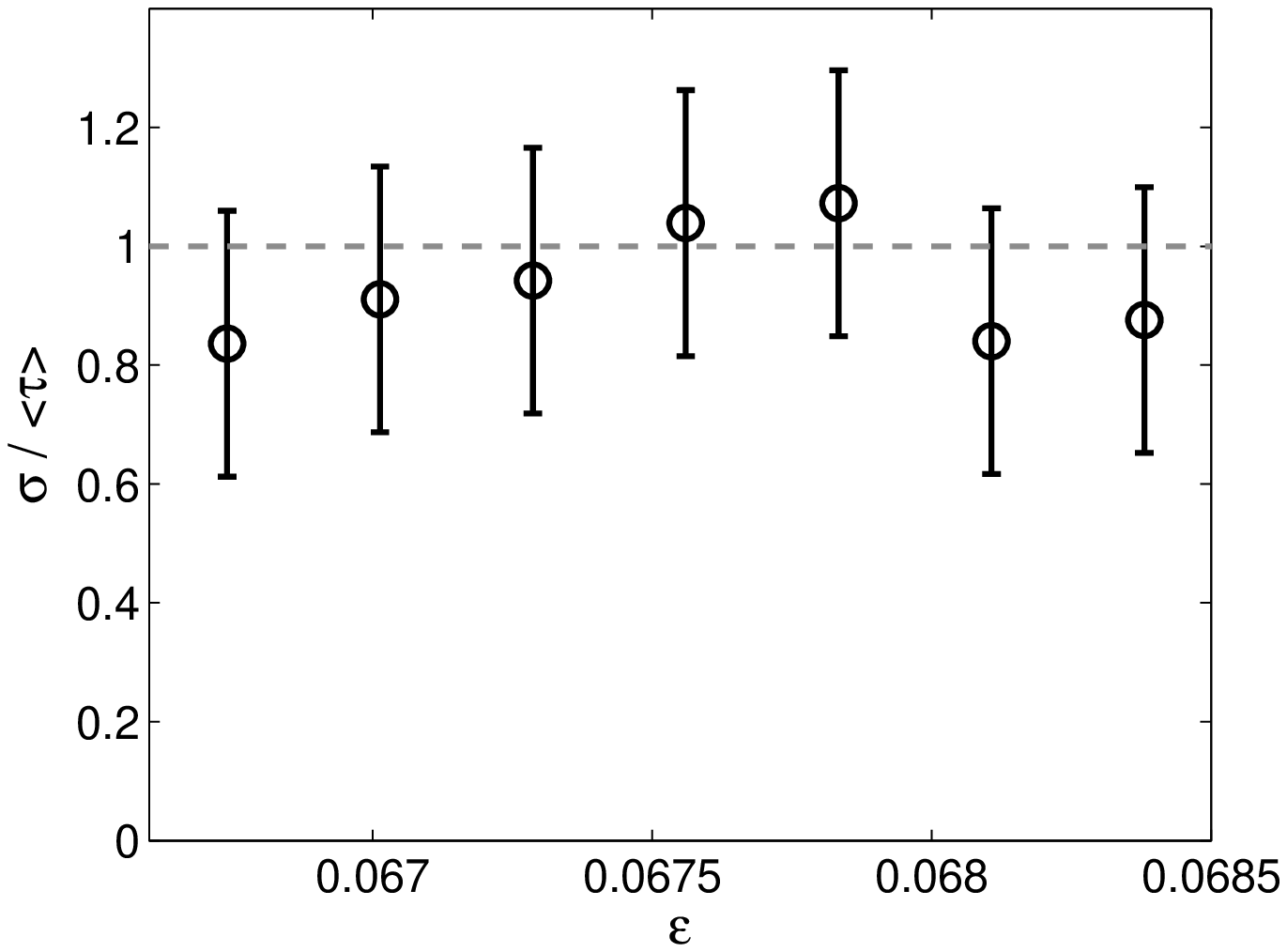}
\caption{Left: distribution function of the lifetime of the turbulence for $\epsilon=6.7 \times 10^{-2}$ and $Re=5.65 \times 10^5$ with $P(\tau)=C \exp(- \tau/\tau_0)$ with $\tau_0=288 \pm 30s$. Each point corresponds to an interval of $150$ s.  Right: ratio of the standard deviation $\sigma_\tau$ over the average $\langle \tau \rangle$.  }
\label{fig_pdf}
\end{center}
\end{figure}

We have performed a large number of measurements ($N=125)$ for $\epsilon=6.7 \times 10^{-2}$ in order to calculate the probability distribution function $\rho(\tau)$ (left Fig. \ref{fig_pdf}). The distribution   displays an exponential tail with $\rho(\tau) \sim \exp( -\tau/\tau_0)$  and  $\tau_0 =288 \pm 30$ s or $1.1 \times 10^{3}$ rotation rate periods (left Fig. \ref{fig_pdf}), which is characteristic  of a memoryless process. The value $\tau_0$ is close to the average lifetime  $ \langle \tau \rangle=267$ s, directly calculated from the set of durations $\tau$.   

\subsubsection{Average lifetime}

\begin{figure}[htb!]
\begin{center}
\includegraphics[width=95mm,height=60mm]{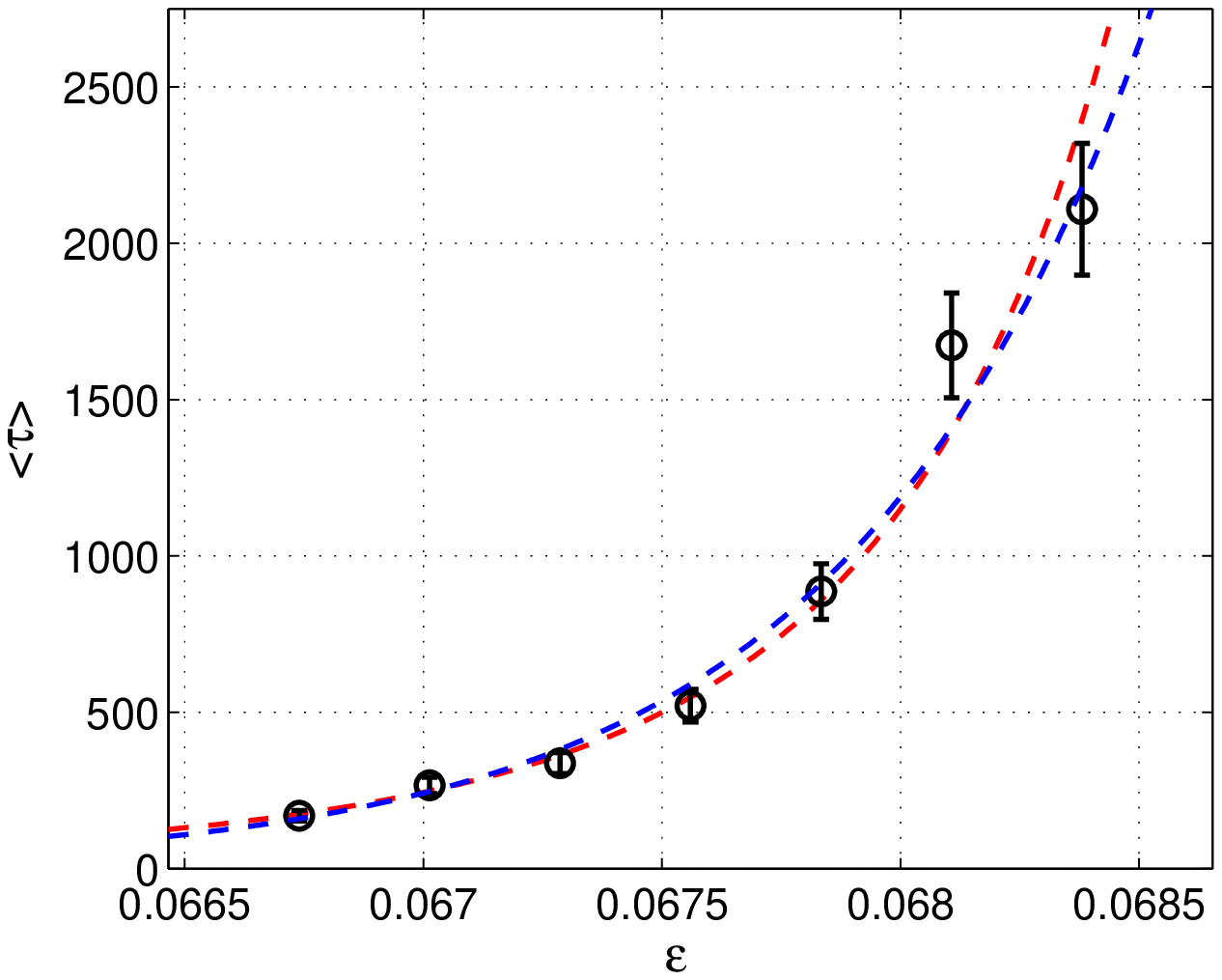}
\includegraphics[width=65mm,height=60mm]{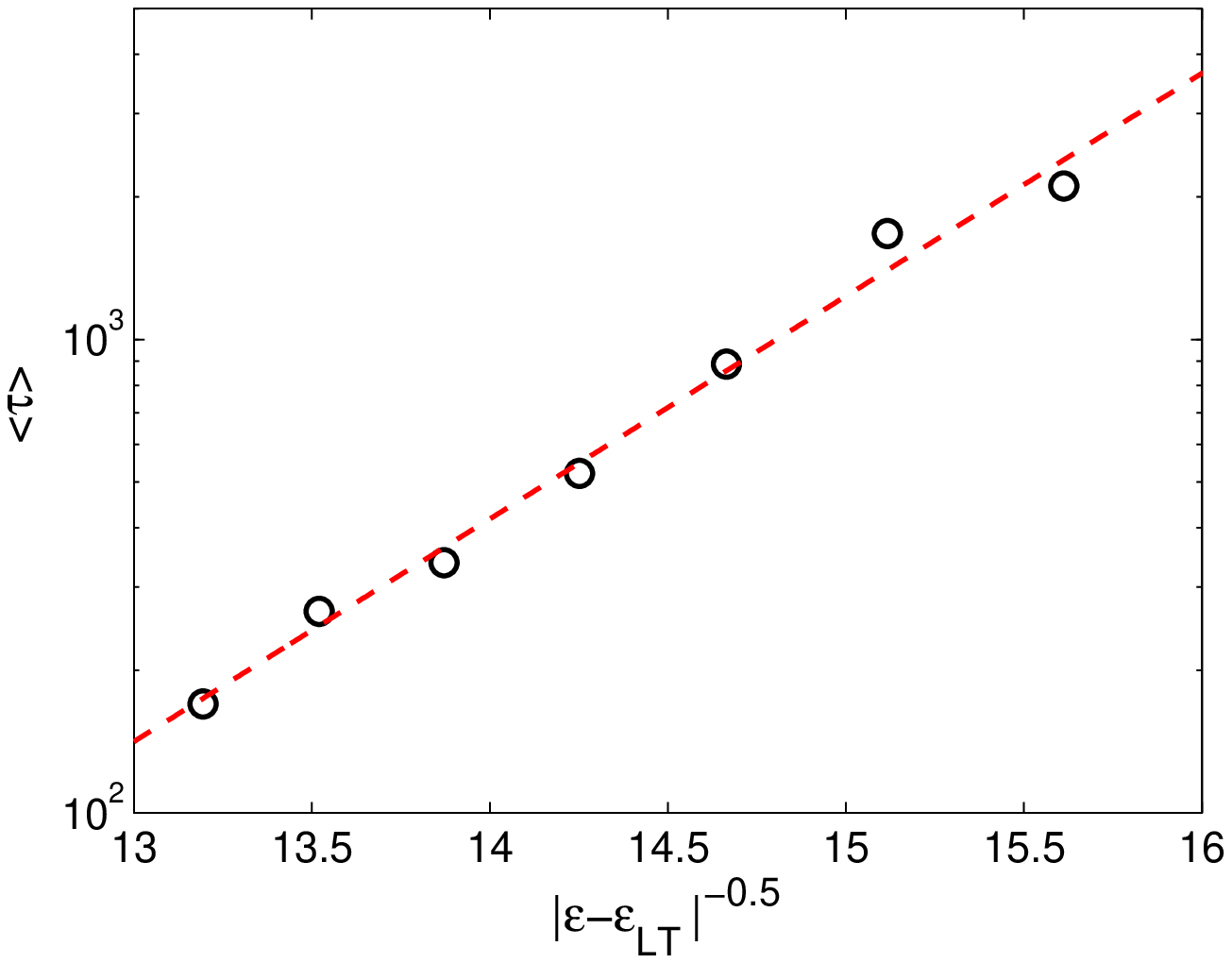}
\caption{Left: Average lifetime of the transient turbulence for $Re=5.65 \times 10^5$ for different $\epsilon<\epsilon_{LT}$ ($\epsilon_{LT}=7.3\times 10^{-2}$). The red curve represents the functional form $\langle \tau \rangle_{th}=   \exp(  a  \vert \epsilon-\epsilon_{LT} \vert ^{-\beta}+b)$ and the blue curve corresponds to the exponential function  $\langle \tau \rangle_{th}=   \exp(  a \epsilon+b)$. Right: Average lifetime $\langle \tau \rangle $ in function of $ \vert \epsilon-\epsilon_{LT} \vert ^{-0.5} $.}
\label{fig_mean_wait}
\end{center}
\end{figure}

We have performed a series of measurements to calculate the average lifetime in the turbulent branch     for   $\epsilon $ between  $6.67 \times 10^{-2}$ and $6.84 \times 10^{-2}$ (Fig. \ref{fig_mean_wait} ). The average is calculated for at least $30-40$ realizations for each precession ratio. The average lifetime $\langle \tau \rangle$ varies from $170$   to $2110$ seconds. It exhibits  a steep increase when $\epsilon$ comes  closer to $\epsilon_{LT}$.

By assuming  that $\langle \tau \rangle$ becomes infinite  when $\epsilon$ is equal to $\epsilon_{LT}$, it is tempting to model the behaviour of the average lifetime $\langle \tau \rangle$   by a critical behaviour with   $\langle \tau \rangle_{th} \sim \vert \epsilon-\epsilon_{LT} \vert ^{-\beta}$. Here, the critical point is  "a priori" defined as equal to $\epsilon_{LT}$ but it cannot be ruled out that the critical point differs from this value. The exponent $\beta$ estimated from such a fit turns out to be larger than $10$, suggesting that the divergence of $\langle \tau \rangle$ is  stronger than a power law. The behaviour of  $\langle \tau \rangle$ is better approximated by an exponential growth with  

\begin{equation}
\langle \tau \rangle_{th}=  \exp( a \vert \epsilon-\epsilon_{LT} \vert ^{-\beta}+b)
\label{supertransient}
\end{equation}
 
\noindent and $(a,b,\beta)=( 1.08,-9.13,1/2)$. The fit is represented by a red curve on the left hand side of Fig. \ref{fig_mean_wait}. Equation (\ref{supertransient}) is similar to the Arrhenius equation and may refer to a behaviour called "supertransient" \cite{grebogi1985super,tel2008chaotic} in dynamical systems.

We have represented the escape rate $ \langle \tau \rangle $ as a function of $ \vert \epsilon-\epsilon_{LT} \vert ^{-0.5} $ in a  logarithm  scale for the ordinate (right Fig.\ref{fig_mean_wait}). The circles are   aligned along the red line, justifying the functional form chosen. The coefficients  $a$ and $b$ have been estimated from the linear relation   $\log(\langle \tau \rangle)=a \vert \epsilon-\epsilon_{LT} \vert ^{-0.5} +b$  with a  least square method, which gives  a smaller relative error (cf. equation \ref{rel_error}).  

Actually, the value $\beta$ can  be varied in the range $[0.1,0.7]$  without changing significantly the quality of the fit. We define the relative error function $r_e$ as

\begin{equation}
 r_e= \left( \frac{1}{N} \sum ^N \frac{(\langle \tau \rangle -\langle \tau \rangle_{th})^2 }{\langle \tau \rangle_{th}^2 } \right)^{\frac 1 2} 
 \label{rel_error}
\end{equation}

\noindent with $N=7$ the number of experimental points. The relative error $r_e$ is between $ 9.45 \%$ and $9.55 \%$ 
for $\beta \in [0.1,0.7]$ and all the functions $\langle \tau \rangle_{th}$ collapse almost on the same curve.  The main contribution to the relative error  $r$ comes from the mean durations $\langle \tau \rangle$ at $\epsilon$ larger than $6.8\times 10^{-2}$. In order to calculate more precisely the exponent $\beta$, measurements at larger $\epsilon$, i.e. with much larger $\langle \tau \rangle$, would be needed. The quality of the fit depends also on the value of the  critical point, here defined "a priori" as $\epsilon_{LT}$.

 We point out that the behaviour of $\langle \tau \rangle$ can be also approximated by an exponential function with $\langle \tau \rangle_{th} \sim  \exp( a \epsilon)$ with a relative error $r_e= 7\%$ (Right Fig.\ref{fig_mean_wait}, blue dashed line). The fit is even better than with the equation \ref{supertransient}, but it implies that the turbulence remains always metastable. %We remark that a Taylor expansion of $\vert \epsilon-\epsilon_{LT} \vert ^{-\beta}$ for $\epsilon$  smaller than $\epsilon_{LT}$, shows that the two previous functional forms of  $\langle \tau \rangle_{th} $ (exponential and the equation \ref{supertransient}) are equivalent in our range of $\epsilon$. 

{     The difference between  the relative error of the exponential law and the equation \ref{supertransient} comes from the contribution of the two last points, where the incertitude on the measured $\langle \tau \rangle$ is large (around $10\%$). It implies that our limited set of measurements does not allow us to decide if there is a threshold (cf equation \ref{supertransient})  or if the turbulence remains metastable for larger $\epsilon$ (exponential law). An estimation of $\left<\tau\right>$ for $\epsilon_{LT}$ with the exponential law gives $\langle \tau \rangle=10^6$ s or $11$ days. It implies that the experimental investigation of $\langle \tau \rangle$ near $\epsilon_{LT}$ is impossible with our set-up. }

With the same set of measurements, we have calculated the standard deviation of $\langle  \tau \rangle$ for different  $\epsilon$. The  standard deviation $\sigma_{\tau}$ normalized  by $\langle \tau \rangle$ (right Fig.\ref{fig_pdf}) is  close to one, as expected for an exponential distribution.

%---------------------------------------------------
\subsection{Relaxation process}
\label{relaxation_sec}
%---------------------------------------------------

\begin{figure}[htb!]
\begin{center}
\includegraphics[width=80mm,height=50mm]{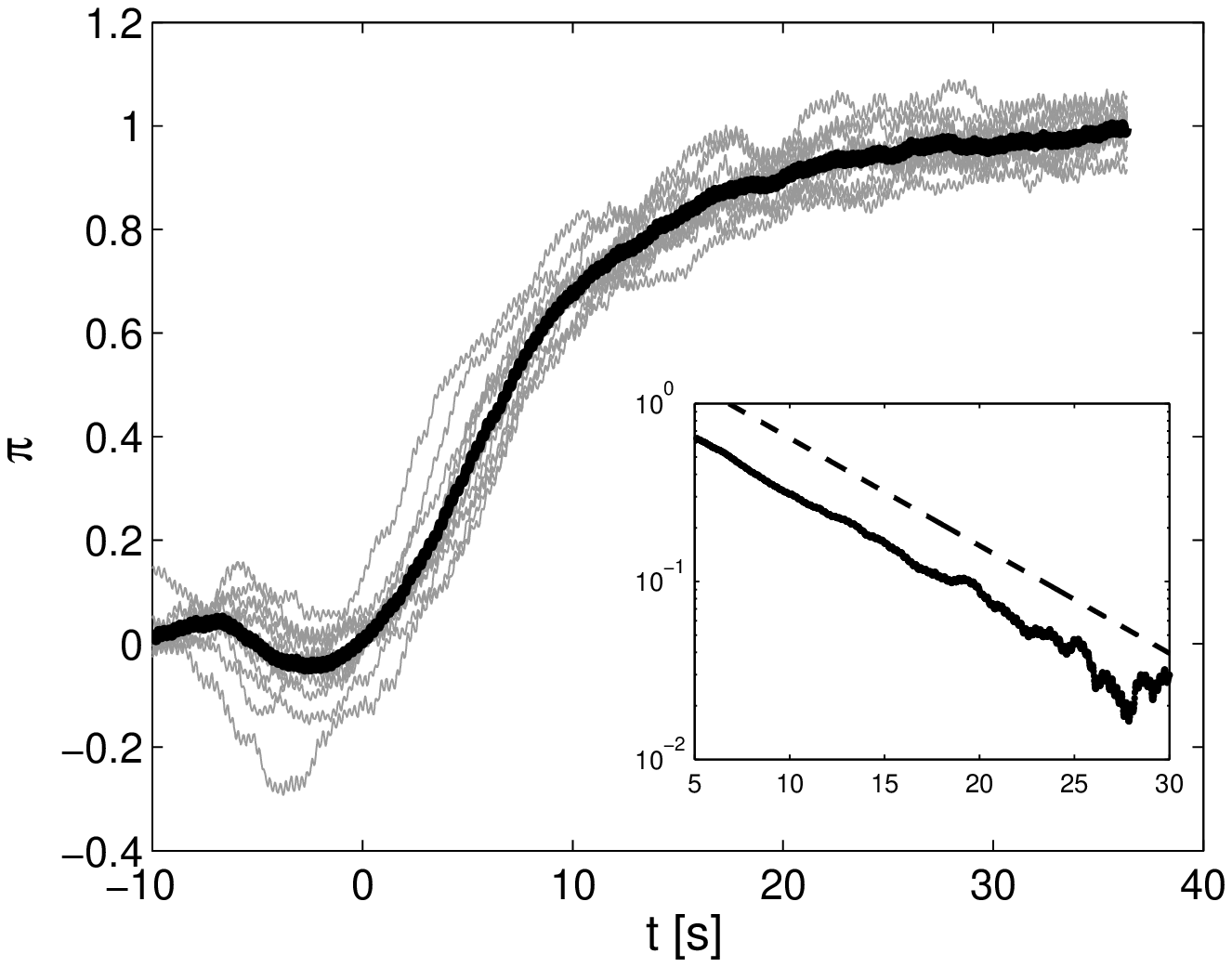}
\includegraphics[width=80mm,height=50mm]{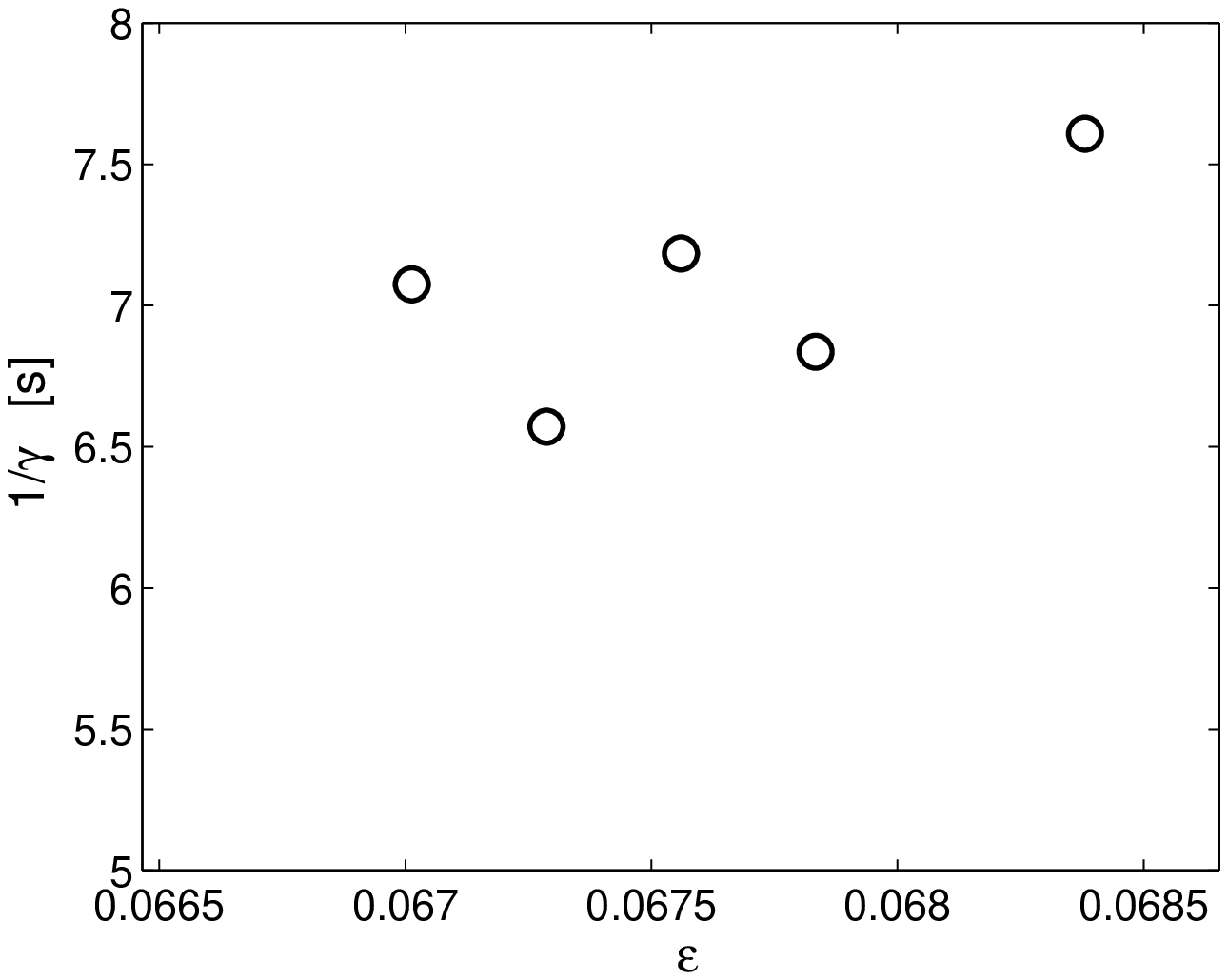}
\caption{Left : superposition of different realizations (in grey) of the relaxation process for  $Re=5.65 \times 10^5$ and $\epsilon=6.7\times 10^{-2}$ of the low-pass filtered pressure. The black curve is the average, which follows an exponential function (insert). Right: the relaxation time $\gamma^{-1}$ versus the precession ratio.}  
\label{fig_relaxation}
\end{center}
\end{figure}

The relaxation from the turbulent to the lower branch is a fast process, compared to the transient lifetime of turbulence. We noticed by direct visualization that the turbulent flow still fills the entire vessel  (like in the right hand side of Fig.\ref{fig_regimes}) before the collapse of the turbulence. We have reported different realizations (in grey) of the relaxation of the pressure $\pi$ on the left of Fig. \ref{fig_relaxation} for  $Re=5.65 \times 10^5$ and $\epsilon=6.7\times 10^{-2}$. The  procedure is identical to the one detailed  in section \ref{transition_turb}.

All the realizations are concentrated around the average (left hand side of Fig.\ref{fig_relaxation}, black curve). Unlike the stochastic process of the lifetimes, the relaxation seems to be deterministic with a small variability in the realizations.  The transition lasts roughly $20$ seconds. The end of the process ($t>5$s) follows an exponential relaxation with $\pi-1 \sim \exp(-\gamma t)$ and $\gamma>0$  (insert in left hand side Fig.\ref{fig_relaxation}). The relaxation times  $\gamma^{-1}$ are reported    for different $\epsilon$ on the right of Fig. \ref{fig_relaxation}. The parameter  $\gamma^{-1}$ is almost constant and stays in the range $[6.5,7.5]$ seconds. For $\epsilon=6\times 10^{-2}$  ($<\epsilon_{TL}$), the relaxation remains also exponential with a similar time scale $\gamma^{-1}=7.69$s, even if the turbulent state is no more metastable in this range of $\epsilon$ (cf section \ref{metastability}). 
 
%--------------------------------------------------------------------------------------------------------------------

\section{Conclusion}
\label{Discussion}
%--------------------------------------------------------------------------------------------------------------------

\subsection{Summary}

In the present study of a precession driven flow, we have confirmed the existence of a subcritical bifurcation  to turbulence  associated with a discontinuity and a hysteresis. The observed transition to turbulence is a robust process  and occurs above a well defined threshold  $\epsilon_{LT}$, different from the one of the parametric instability of the forced Kelvin mode. Both the pressure and power consumption measurements exhibit an abrupt variation at the transition from and to turbulence. The behaviour of the power consumption changes at the threshold, with a linear dependence of the power on the precession ratio. The pressure measurements suggest  that the transition is associated with a slowdown of the solid body rotation. The threshold $\epsilon_{LT}$   of the transition depends weakly on the Reynolds numbers. 

In the   region of the hysteresis $\epsilon \in[\epsilon_{TL},\epsilon_{LT}]$, the turbulence can suddenly collapse and lead to a definitive relaminarization of the flow. For  a given precession ratio and Reynolds number, the   lifetimes $\tau$ of the turbulence are random and are exponentially distributed . The average lifetimes $\langle \tau \rangle$ of the turbulence increase rapidly when $\epsilon$ approaches the threshold $\epsilon_{LT}$ of the sustained turbulence. The growth of the lifetimes $\tau$ is well modelled either by the functional form $\langle \tau \rangle   \sim \exp(a \vert \epsilon-\epsilon_{LT} \vert^{-\beta})$ with $a$ and $\beta$ positive {    or an exponential law}.  A detailed study of the relaxation process shows that the slow component of the pressure follows an exponential function with a relaxation rate weakly dependent of $\epsilon$, even for $\epsilon<\epsilon_{TL}$.

\begin{figure}[htb!]
\begin{center}
\includegraphics[width=120mm,height=75mm]{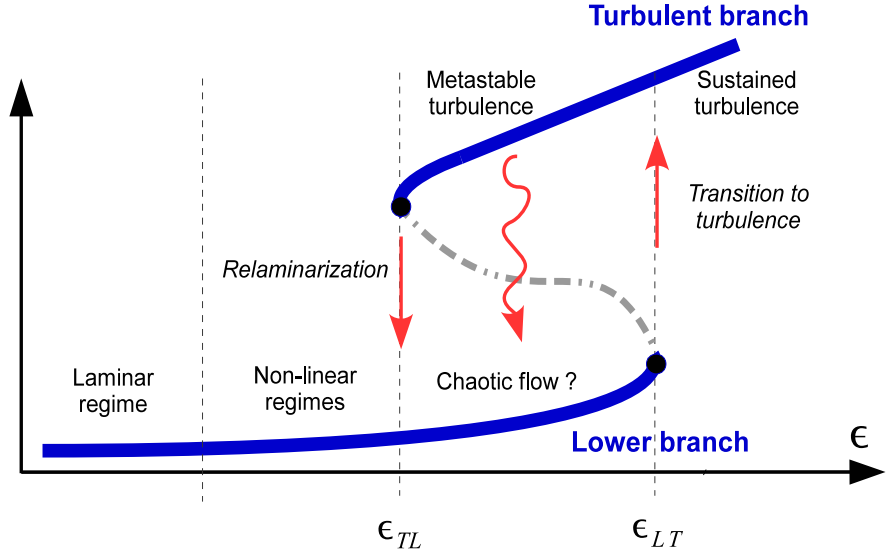} 
\caption{	Schematic bifurcation diagram of the precessing cylinder for $\epsilon< 16 \times 10^{-2}$ and $Re \in [10^5,10^6]$. 
}  
\label{fig_sumup}
\end{center}
\end{figure}

As a conclusion, a schematic bifurcation diagram is  presented on the Fig. \ref{fig_sumup}. Three regimes are observed : the laminar and the non-linear regimes on the $L$ branch and the turbulent regime  on the $T$ branch (cf section \ref{physical_reg}). The $L$ branch becomes unstable at $\epsilon_{LT}$ and the system undergoes a subcritical bifurcation to turbulence. The flow comes back to the $L$ branch for $\epsilon$ smaller than $\epsilon_{TL}$ (cf section \ref{subcritical}). Between $\epsilon_{TL}$ and $\epsilon_{LT}$, the turbulent regime is metastable and exhibits transient lifetimes (cf section \ref{section_life_time}). {    The sneaking arrow corresponds to the relaxation process to the lower branch}. The grey line represents the frontier between the basins of attraction of each branch, even if the turbulence is not  strictly speaking an attractor. The collisions between the grey and blue lines (black dots) are only symbolic and do not refer to a saddle-node bifurcation.

%-------------------------------------------------------------------------------------------
\subsection{Discussion} 
%-------------------------------------------------------------------------------------------

The observation of a  well-defined threshold $\epsilon_{LT}$ for a sustained turbulence  may lead to  different conclusions. It could indicate  the presence of a linear instability past the onset $\epsilon_{LT}$. The instability could also be triggered by non-normal and non-linear mechanisms. Moreover, if we consider the flow as already chaotic before the instability, the transition could be explained by a crisis\cite{grebogi1982chaotic} of  this chaotic set. This bifurcation refers to the collision of the basin of attraction of two  chaotic attractors, leading to the emergence of only one attractor. However, a non-linear instability is not consistent with  the measured robustness of the threshold, whereas the crisis would be associated with random waiting times before the transition. The linear instability is thus the best candidate for the transition to turbulence. As the instability of the laminar flow occurs for smaller precession ratios than $\epsilon_{LT}$, the mechanism of this new instability could differ from the parametric instability of the Kelvin modes \cite{kerswell1999secondary}.   
 
\begin{acknowledgments}
The authors are grateful for support by the Helmholtz Allianz LIMTECH and kindly acknowledge discussions with T. Weier, B. Wustmann , T. Albrecht, C. Nore, J. L\'eorat, A. Tilgner and J. Noir.   
\end{acknowledgments}

\bibliography{references}

\end{document}